\documentclass[10pt]{article}
\usepackage{amsfonts}
\usepackage{amsmath,amssymb}
\usepackage{graphicx}
\usepackage{cite}
\newcommand{\abst}{~~}

\newcommand{\bra}[1]{\langle{#1}|}
\newcommand{\braket}[1]{\langle{#1}\rangle}
\newcommand{\ket}[1]{|{#1}\rangle}
\newcommand{\degK}{^\circ\mathrm{K}~}
\newcommand{\mbo}[1]{$#1$}

\newcommand{\mpl}{M_{\rm Pl}}

\newcommand{\power}[1]{\times 10^{#1}}

\newcommand{\Ba}{\begin{eqnarray}}
\newcommand{\Ea}{\end{eqnarray}}
\newcommand{\bea}{\begin{eqnarray}}
\newcommand{\eea}{\end{eqnarray}}
\newcommand{\ba}{\begin{eqnarray*}}
\newcommand{\ea}{\end{eqnarray*}}

\newcommand{\nn}{\nonumber}

\newcommand{\eps}{\varepsilon}

\newcommand{\gv}{\mbox{GeV}}
\newcommand{\mv}{\mbox{MeV}}
\newcommand{\epo}{\,.}

\newcommand{\cs}{\,,}
\newcommand{\semis}{\,;\;\;}
\newcommand{\comas}{\,,\;\;}
\newcommand{\be}{\begin{equation}}
\newcommand{\ee}{\end{equation}}

\newcommand{\MSb}{$\overline{\rm MS}$ }

\newcommand{\cL}{{\cal L}}

\newcommand{\D}{\mathrm{d}}
\newcommand{\E}{\mathrm{e}}
\newcommand{\I}{\mathrm{i}}

\newcommand{\vev}[1]{\langle0| #1 |0\rangle}
\newcommand{\tpl}{t_{\rm Pl}}

\newcommand{\gapprox}{\raisebox{-.2ex}{$\stackrel{\textstyle>}{\raisebox{-.6ex}[0ex][0ex]{$\sim$}}$}}

\newcommand{\gmunu}{g_{\mu \nu}}
\newcommand{\munu}{{\mu \nu}}

\newcommand{\rosi}{{\rho \sigma}}

\newcommand{\gmunuc}{g^{\mu\nu}}

\newcommand{\pamu}{\partial_\mu}
\newcommand{\panu}{\partial_\nu}
\newcommand{\paro}{\partial_\rho}
\newcommand{\pasi}{\partial_\sigma}
\newcommand{\ha}{\frac12}

\newcommand{\np}{{\em Nucl.\ Phys.\ }{\bf B}}
\newcommand{\pl}{{\em Phys.\ Lett.\ }}

\newcommand{\prl}{{\em Phys.\ Rev.\ Lett.\ } }
\newcommand{\pr}{{\em Phys.\ Rev.\ }}

\begin{document}
\title{%
\vskip-3.5cm{\baselineskip14pt
\centerline{\hfill\footnotesize DESY~14-013,~~HU-EP-14/05}
}
\vskip1.5cm
Higgs inflation and the cosmological constant
}

\author{
{\sc Fred Jegerlehner},
\\
\\
{\normalsize Humboldt-Universit\"at zu Berlin, Institut f\"ur Physik,}\\
{\normalsize  Newtonstrasse 15, D-12489 Berlin, Germany}\\
{\normalsize Deutsches Elektronen-Synchrotron (DESY),}\\
{\normalsize Platanenallee 6, D-15738 Zeuthen, Germany}
}

\date{}

\maketitle


\begin{abstract}
We discuss the impact of the Higgs discovery and its revealing a very
peculiar value for the Higgs mass. It turns out that the Higgs not
only induces the masses of all SM particles, the Higgs, given its
special mass value, is the natural candidate for the inflaton and in
fact is ruling the evolution of the early universe, by providing the
necessary dark energy which remains the dominant energy density. In a
previous paper I have shown that running couplings not only allow us
to extrapolate SM physics up to the Planck scale, but equally
important they are triggering the Higgs mechanism when the universe
cools down to lower temperatures. This is possible by the fact that
the bare mass term in the Higgs potential changes sign at about
$\mu_0\simeq 1.4\power{16}~\gv$ and in the symmetric phase is enhanced
by quadratic terms in the Planck mass.  Such a huge Higgs mass term is
able to play a key role in triggering inflation in the early
universe. In this article we extend our previous investigation by
working out the details of a Higgs inflation scenario. We show how
different terms contributing to the Higgs Lagrangian are affecting
inflation. Given the SM and its extrapolation to scales $\mu>\mu_0$ we
find a calculable cosmological constant $V(0)$ which is weakly scale
dependent and actually remains large during inflation. This is
different to the Higgs fluctuation field dependent $\Delta V(\phi)$,
which decays exponentially during inflation, and actually would not
provide a sufficient amount of inflation to solve the CMB horizon
problem. The fluctuation field has a different effective mass which
shifts the bare Higgs transition point to a lower value $\mu'_0\simeq 7.7
\power{14}~\gv\,.$ We also show that for SM inflation standard slow-roll
inflation criteria are obsolete. Miraculously, the huge difference
between bare and renormalized cosmological constant is nullified
either by the running of the SM couplings or by vacuum rearrangement
somewhat before the Higgs phase transition takes place. This solves
the notorious cosmological constant problem.  Like in the case of the
standard hierarchy problem concerning the quadratic divergences, also
the quartically divergent vacuum energy exhibits a coefficient
function which exhibits a zero very close to the zero of the quadratic
coefficient function. While the Higgs today is only talking very
weakly to the rest of the world, in the early universe it was
all-dominating and shaping the universe to look as we see it
today. The role of the Higgs in reheating and baryogenesis is
emphasized. SM inflation implies reheating by production of
top--anti-top pairs predominantly. In our scenario inflation and the
EW phase transition happen not far below the Planck scale where baryon
and lepton number violating dimension 6 operators, suppressed by at
most $(\mu/\mpl)^2>(\mu_0/\mpl)^2=10^{-6}$, provide sufficient $B$ and
$L$ violation to trigger baryogenesis. The heavy ``charged'' Higgs
decays $H^+\to t\bar{d},u\bar{b}$ and $H^-\to d\bar{t},b\bar{u}$,
proportional to the CP-violating CKM elements $V_{td}$ and $V_{bu}$ or
their conjugates, together with the fact that the Higgs transition and
its closely following EW phase transition is driving the
system out of equilibrium, provide the proper condition for
baryogenesis to derive from SM physics. After the EW phase transition
the now heavy flavors decay into the lighter ones ending up as normal
matter.
\medskip

\end{abstract}

\section{Introduction}
The discovery of the Higgs by ATLAS~\cite{ATLAS} and CMS~\cite{CMS} at
the LHC, together with the fact that new physics still did not show
up, already has changed the paradigm about the path to physics at the
high energy scale.  The Standard Model (SM) of particle
physics~\cite{SM,QCD} appears to be finalized by the Higgs
sector~\cite{Higgs}, as it has been required by the theory of mass
generation, and all the main SM parameters are known rather accurately
by now~\cite{pdg}.  In Ref.~\cite{Jegerlehner:2013cta} we have
studied the impact of this new situation:

\abst \textbf{1)} The present status of the SM strengthens the status of the SM as a
low energy effective (LEESM) theory of a cutoff system residing at the Planck
scale, with the Planck mass $\mpl\simeq1.22\power{19}~\gv$ as a
cutoff.  It renders the now finite relationship between bare and
renormalized parameters to have a precise physical meaning. We thus
can calculate the parameters of the bare system residing at the Planck
scale.

\abst \textbf{2)} The SM very likely, within present input parameter uncertainties, 
remains a self-consistent QFT in the perturbative regime up to very
close to the Planck scale, with a stable Higgs vacuum.

\abst \textbf{3)} The quadratically enhanced Higgs potential mass counterterm
has a known scale dependent coefficient, which changes sign at about
$\mu_0\simeq 1.4\power{16}~\gv$. The sign-flip is triggering the Higgs
mechanism, which means that the SM in the early universe has been in
the symmetric phase with four physical very heavy Higgses, while all other SM
particles are essentially massless. At the Higgs transition point the
difference between bare and renormalized masses is nullified, such
that the bare short distance world before the phase transition matches
the renormalized low energy world after the phase transition.  

\abst \textbf{4)} Before the universe has cooled down to undergo the Higgs phase 
transition, the Higgs is triggering inflation and provides the
necessary large dark-energy term corresponding to a large bare cosmological
constant.

In Ref.~\cite{Jegerlehner:2013nna} we emphasized that the SM in the
broken phase has no hierarchy problem. In contrast, in the unbroken
phase (in the early universe) the quadratic enhancement of the mass
term in the Higgs potential is what promotes the Higgs to be the
inflaton scalar field. Thus the ``quadratic divergences'' provide the
necessary condition for the explanation of the inflation profile as
extracted from Cosmic Microwave Background (CMB)
data~\cite{PlanckResults}.

Note that in the unbroken phase, which exists from the Planck scale down to the
Higgs transition not very far below the Planck scale, the bare theory
is the physical one and a hierarchy or fine-tuning problem is not an
issue there.

Standard Model Higgs vacuum stability bounds have been studied some
time ago in Ref.~\cite{Hambye:1996wb,Holthausen:2011aa}, for
example. Surprisingly, the Higgs mass determined by the LHC
experiments revealed a value which just matched or very closely
matched expectations from vacuum stability bounds.  This has then been
elaborated in a number of
papers~\cite{Yukawa:3,degrassi,Moch12,Mihaila:2012fm,Chetyrkin:2012rz,Masina:2012tz,Bednyakov:2012rb,Bednyakov:2012en,Chetyrkin:2013wya,Bednyakov:2013eba,Buttazzo:2013uya,Bednyakov:2013cpa}.
While a majority of (not independent) analyses just find sharply
missing vacuum stability, which means that new physics must be there
to remedy the unstable situation, some analyses obtain results
compatible with vacuum stability up to the Planck
scale~\cite{Jegerlehner:2012kn,Jegerlehner:2013cta,Bednyakov:2013cpa}. The
main issue is the precise value of the top-quark Yukawa coupling,
which we find slightly smaller than some other analyses. Differences
may be related to issues concerning the mass definition of the
top-quark~\cite{Juste:2013dsa} and the proper evaluation of the
on-shell versus \MSb matching conditions as analyzed e.g. in
Ref.~\cite{Jegerlehner:2012kn} or more recently in
Ref.~\cite{Buttazzo:2013uya}. Our study is based on input values
listed in Table~\ref{tab:params}, which also lists corresponding input
values obtained in Ref.~\cite{Buttazzo:2013uya}. Stable vacuum
solutions have been found independently in
Ref.~\cite{Bednyakov:2013cpa}.
\begin{table}[h]
\caption{\MSb parameters at various scales for $M_H=126~\gv$ and
$\mu_0\simeq 1.4\power{16}~\gv$. $C_1$ and $C_2$ are the one- and
two-loop coefficients of the quadratic divergence Eqs.~(\ref{coefC1}) and (\ref{coefC2}),
respectively. The last two columns show corresponding results from
Ref.~\cite{Buttazzo:2013uya}.\smallskip}  {\small \begin{tabular}{ccccc||cc}\hline
\hline\noalign{\smallskip}
coupling $\backslash$ scale  & $M_Z$ & $M_t$ & $\mu_0$ & $\mpl$ & $M_t$~[\citen{Buttazzo:2013uya}] &
$\mpl$~[\citen{Buttazzo:2013uya}] \\  \hline
$g_3$ &   $1.2200$ & $1.1644$ & $0.5271$ & $0.4886$ & 1.1644 &\hphantom{-} 0.4873 \\
$g_2$ &   $0.6530$ & $0.6496$ & $0.5249$ & $0.5068$ & 0.6483 &\hphantom{-} 0.5057 \\
$g_1$ &   $0.3497$ & $0.3509$ & $0.4333$ & $0.4589$ & 0.3587 &\hphantom{-} 0.4777 \\
$y_t$ &   $0.9347$ & $0.9002$ & $0.3872$ & $0.3510$ & 0.9399 &\hphantom{-} 0.3823 \\
$y_b$ &   $0.0238$ & $0.0227$ & $0.0082$ & $0.0074$ && \\
$y_\tau$ &$0.0104$ & $0.0104$ & $0.0097$ & $0.0094$ && \\
$\sqrt{\lambda}$&$0.8983$ & $0.8586$ & $0.3732$ & $0.3749$ & 0.8733  & $\I\:\,$0.1131 \\
$\lambda       $&$0.8070$ & $0.7373$ & $0.1393$ & $0.1405$ & 0.7626  &
- 0.0128         \\
$C_1$ & $-6.768$ & $-6.110$ & $\hphantom{0}0\hphantom{.0000}$ & $0.2741$ & $$ & $$\\
$C_2$ & $-6.672$ & $-6.217$ & $\hphantom{0}0\hphantom{.0000}$ & $0.2845$ & $$ & $$\\
$m[GeV]$&$89.096$ & $89.889$ & $97.278$ & $96.498$ &97.278&  \\ \hline
\end{tabular}
\label{tab:params}}
\end{table}

In a LEESM scenario the only quadratic divergences show up in the
renormalization of the mass  term in the Higgs potential 
\bea
m_0^2=m^2+\delta m^2\;;\;\; \delta m^2= \frac{\Lambda^2}{32 \pi^2}\,C
\label{barem2}
\eea
which communicates the relationship between the bare $m_0$ and the
renormalized mass $m$. The one-loop coefficient function~\cite{Veltman:1980mj}
$C_1$ may be written as
\bea
C_1=2\,\lambda+\frac32\, {g'}^{2}+\frac92\,g^2-12\,y_t^2\,,
\label{coefC1}
\eea
and is uniquely determined by dimensionless couplings. Surprisingly,
taking into account the running of the SM couplings, which are not
affected by quadratic divergences such that standard RG equations
apply, the coefficient of the quadratic divergence of the Higgs mass
counterterm exhibits a zero. This has been emphasized in
Ref.~\cite{Hamada:2012bp}. The next-order correction, first calculated
in Refs.~\cite{Alsarhi:1991ji,Jones:2013aua},
\bea
C_2&=&C_1+ \frac{\ln (2^6/3^3)}{16\pi^2}\, [
18\,y_t^4+y_t^2\,(-\frac{7}{6}\,{g'}^2+\frac{9}{2}\,g^2
             -32\,g_s^2) \nn \\
             &&-\frac{87}{8}\,{g'}^4-\frac{63}{8}\,g^4 -\frac{15}{4}\,g^2{g'}^2
             +\lambda\,(-6\,y_t^2+{g'}^2+3\,g^2)
             -\frac{2}{3}\,\lambda^2]\,,
\label{coefC2}
\eea
numerically does not change significantly the one-loop result. For
$M_H=126~\gv$, and given our set of \MSb input parameters at the
scale $M_Z$, the zero of $C_1$ is at $\mu_0\simeq 1.4\power{16}~\gv$
the one of $C_2$ at $\mu_0\simeq 1.8\power{16}~\gv\,$. For the same
Higgs mass the RG $\beta$-function $\beta_\lambda$ has a zero at
$1.3\power{17}~\gv\,$. Since the difference between $C_1$ and $C_2$
is small, we will adopt $C_1$ and the corresponding value for $\mu_0\,$, in what follows.
For later use we also list the zeros for $C'_i=C_i+\lambda$ and $X_i=\frac18\,\left(2C'_i-\lambda\right)$
(see Eqs.(\ref{coefC1prime}) and (\ref{finetuning}) below) in Table~\ref{tab:zeros}.
\begin{table}[h]
\caption{The location of the zeros of $C_i$, $C'_i=C_i+\lambda$ and $X_i=\frac18\,\left(2\,C'_i-\lambda\right)$
as a function of scale in GeV, at one ($i=1$) and two ($i=2$) loops. Parameters as in
Table~\ref{tab:params}.\smallskip}  {\footnotesize \begin{tabular}{cc|cc|cc}\hline
\hline\noalign{\smallskip}
 $C_1$ & $C_2$ & $C'_1$ & $C'_2$ & $X_1$& $X_2$ \\ \noalign{\smallskip}\hline\noalign{\smallskip}
 $1.42\power{16}$ & $1.82\power{16}$& $7.77\power{14}$ & $9.94\power{14}$ &
$3.25\power{15}$ & $4.15\power{15}$
\\ \noalign{\smallskip}\hline
\end{tabular}
\label{tab:zeros}}
\end{table}

Above the transition point the number of massless degrees of freedom
(radiation) of the SM consists of $g_f=90$ fermionic degrees of
freedom and $g_b=24$ bosonic ones such that the effective number of
degrees of freedom is given by 
\bea g_*(T)=g_b(T)+\frac78\,g_f(T)=102.75\cs
\eea 
the factor $7/8$ accounts for the Pauli exclusion principle which
applies for the fermions. The four Higgses in the symmetric phase
have equal masses, and are very heavy. Highly relativistic particles
contribute to the radiation density
\bea
\rho_{\rm rad}(T)=\frac{\pi^2}{30}\,g_*(T)\,T^4\cs
\eea 
according to the Stefan-Boltzmann law.

\begin{figure}[t]
\centering
\includegraphics[height=5.4cm]{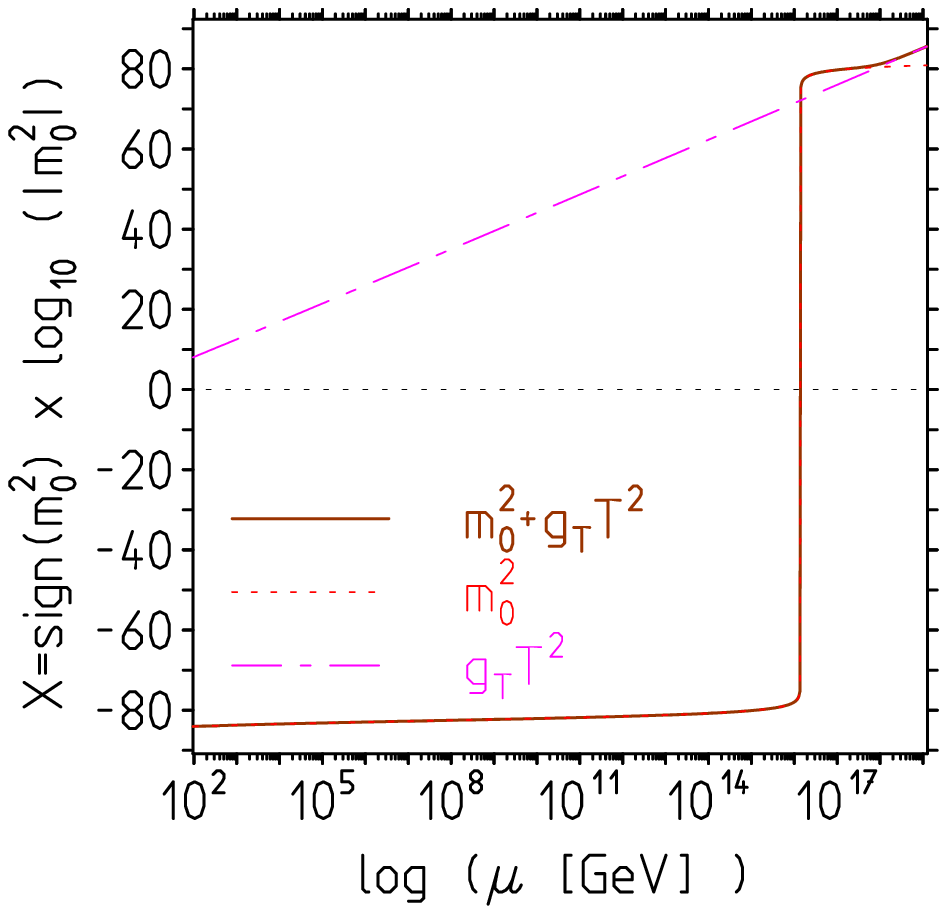}
\includegraphics[height=5.4cm]{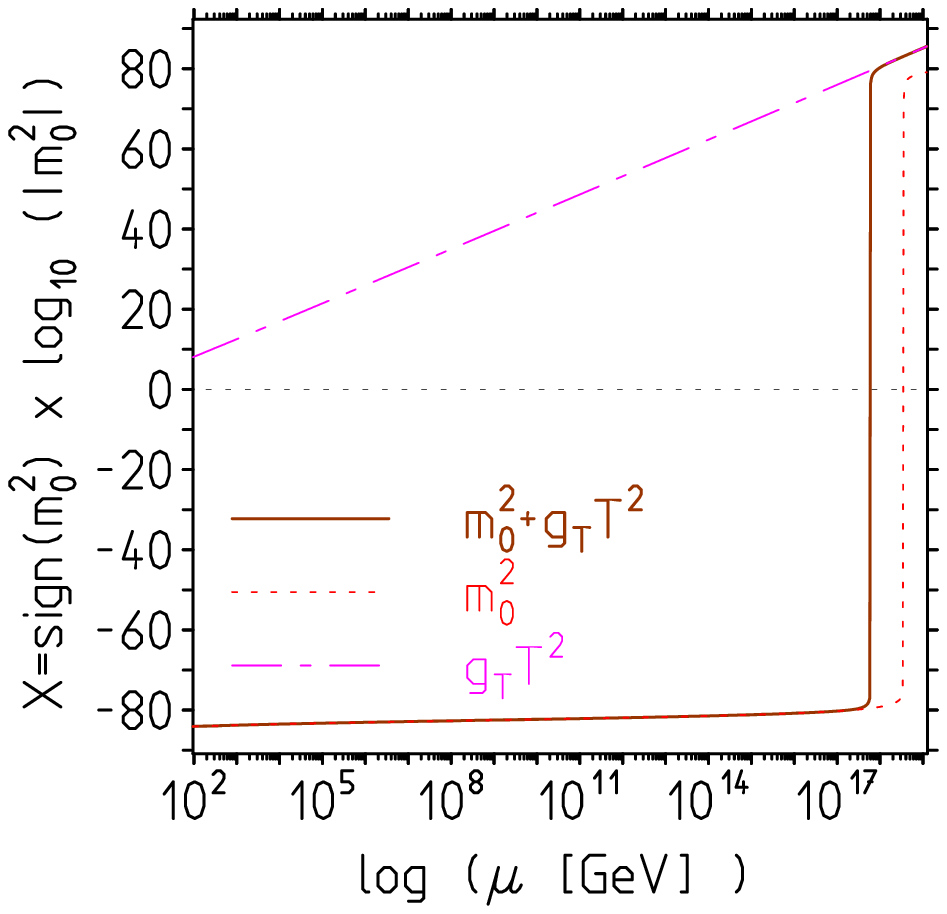}
\caption{The role of the effective bare mass in the finite temperature
SM. Left: for $\mu_0\sim 1.4 \power{16}~\gv$ ($M_H\sim
126~\gv$, $M_t\sim 173.5~\gv$).
Right: finite temperature delayed transition for
$\mu_0\sim 6 \power{17}~\gv$ ($M_H\sim 124~\gv$, $M_t\sim 175~\gv$),
the $m^2_0$ term  alone is flipping at about $\mu_0\sim 3.5 \power{18}~\gv$.
Inflation requires \mbo{m_0^2 > g_T\,T^2}!}
\label{fig:FT}
\end{figure}

In the early hot phase of the universe finite temperature effects play
an important role. They in particular affect the electroweak (EW) phase transition
as well as inflation~\cite{FiniteTemp1,FiniteTemp2,FiniteTemp3,Dine:1992wr}. The
leading effects enter the finite temperature potential 
\bea
V(\phi,T) =
\frac{1}{2}\,\left(g_T\,T^2-\mu_\phi^2\right)\,\phi^2+\frac{\lambda}{24}
\,\phi^4+ \cdots \,. 
\label{FTpotential}
\eea
by a modified effective mass term. 
In the SM $g_T$ is given by~\cite{Dine:1992wr}
\bea
g_T=\frac{1}{16}\,\left[3\,g^2+{g'}^{2}+4\,y_t^2+\frac23\,\lambda
\right]\,,
\eea
and with the couplings given in Table~\ref{tab:params} we estimate 
$g_T \approx 0.0980$ at $\mpl$. RG methods allow us to
calculate the mass parameter $\mu_\phi^2$, which in the symmetric phase, at scales
$\mu>\mu_0$, is given by the symmetric bare mass term
\bea
-\mu_\phi^2\to m_0^2=m^2+\delta m^2>0\,.
\eea
The EW phase transition, requiring $m^2_{\rm eff}= g_T\,T^2 +m_0^2 <0$
can only take place after the Higgs transition $m_0^2 >0 \to m_0^2 <0$,
whereby it is important that $m_0^2$ is large due to the
quadratic enhancement of the mass terms. This implies that the
Higgs transition essentially triggers the EW phase transition to
happen at a very high scale not far below $\mu_0$, as we illustrate in
Fig.~\ref{fig:FT}.
What we call \textit{Higgs transition} corresponds to the would-be EW phase transition
in the zero temperature SM. Standard EW phase transition scenarios  
(see e.g. Refs.~\cite{Dine:1992wr,Buchmuller:1995sf,Kajantie:1996mn,Rubakov:1996vz,Chung:2012vg}
and references therein)  assume $\mu_\phi^2>0$ in the finite
temperature potential (\ref{FTpotential}) to hold up to the Planck scale.

It should be noticed here that in the symmetric phase we actually do
not know the parameter $m^2$ which corresponds to the renormalized mass.
The only possibility to constrain it phenomenologically is via
information we have on inflation. We \textbf{assume} it to be small
relative to the quadratically enhanced $\delta m^2$. Information
extracted from CMB fluctuations on the inflaton mass yielded estimates like
\bea
m_\phi \sim 10^{-6}\,\mpl\cs
\label{mphiguess}
\eea
which essentially supports our assumption.

It is well known that, in principle, the Higgs could be the scalar
inflaton field, which is able to explain the phenomenon of inflation in
the early
universe~\cite{Guth:1980zm,Starobinsky:1980te,Linde:1981mu,Albrecht:1982wi,Mukhanov:1981xt,Mukhanov:1985rz,Mukhanov:1990me}
(see also the recent Refs.~\cite{Bezrukov:2009db,Bezrukov:2012hx} and references
therein). Inflation requires an exponential growth $a(t)\propto
\E^{Ht}$ of the Friedman-Robertson-Walker (FRW) radius $a(t)$ of the
universe, where $H(t)=\dot{a}/a(t)$ is the Hubble constant at cosmic
time $t$. Data strongly supporting the existence of an inflation era
in the evolution of the universe are provided by CMB observations,
most recently from the Planck mission (see Ref.~\cite{PlanckResults}
and references therein). The SM Higgs affects the evolution of the
universe by its contribution to the energy-momentum tensor. Given the
Higgs Lagrangian
\bea
\cL(\phi)=\frac12\,\gmunuc\,\pamu \phi\, \panu \phi-V(\phi)\cs
\eea
its contribution to the symmetric energy-momentum tensor 
\Ba
\Theta^\mu_{\:\:\nu}=\frac{\partial \cL}{\partial (\partial_\mu \phi)}\,\partial_\nu\phi-\delta^\mu_{\:\:\nu}\cL
\Ea
reads
\Ba
\Theta_\munu=\ha\,\pamu \phi\, \panu \phi+\ha\,\left(g^\rosi \paro
\phi\, \pasi \phi\right)\,\gmunu+V(\phi)\,\gmunu \epo
\Ea
In the ground state the gradient terms of \mbo{\phi} do not contribute
and we obtain 
\Ba
T^\mathrm{vac}_\munu\equiv \vev{\Theta_\munu}
=\rho_\mathrm{vac}\,\gmunu=V(\phi_0)\,\gmunu \epo
\Ea
Actually, in the ground state this is the only covariant and
covariantly conserved form possible.  Here, by the Einstein equation
$$R_\munu-\ha\, R\, \gmunu=\kappa T_\munu\cs$$ the Higgs directly
talks to gravity! This is true in any case in the symmetric phase of
the SM where\footnote{At later stages, when long range phenomema come
into play and the SM undergoes spontaneous symmery breaking, other
vacuum condensates participate, besides the Higgs VEV $\braket{H}=v$
generated at the EW phase transiton, in particular the quark and gluon
condensates after the QCD phase transition.}
$$T_{\mu\nu\,\mathrm{SM}}^\mathrm{vac}=-\vev{\cL_\mathrm{SM}}\,g_{\mu\nu}=V(\phi_0)\,\gmunu\epo$$
Indeed the Planck medium factually is unifying gravity with the rest
oft the world, on the one hand through Einstein's equation, on the
other hand since it is Newton's gravitational constant which
determines the intrinsic scale of the Planck medium. The Higgs
contribution $T^\mathrm{vac}_\munu$ is to be identified as a
contribution to the classical ideal fluid energy-momentum tensor
\Ba
T_\munu=\left(\rho+p\right)\,u_\mu u_\nu-p\,\gmunu\cs
\Ea 
where $\rho$ is the density, $p$ the pressure, $u^\mu\equiv \D
x^\mu/\D s$ is the contravariant four velocity of the fluid and
$u_\mu= \gmunu u^\nu$.  The
comparison shows that the Higgs implies a pressure
\Ba
p_\mathrm{vac}=p_\mathrm{\phi}=-\rho_\mathrm{\phi}\epo
\Ea
On the classical field level, assuming spatial isotropy (i.e. $\phi$
only depends on time), the Higgs 
contribution to energy density and pressure is given by
\begin{eqnarray}
\rho_\phi=\frac12\,\dot{\phi}^2+V(\phi)\semis
p_\phi=\frac12\,\dot{\phi}^2-V(\phi)\epo
\end{eqnarray}
The second Friedman equation has the form
\bea
\ddot{a}/a=-\frac{\ell^2}{2}\,\left(\rho+3p\right)\cs
\eea 
where $\ell^2=8\pi G/3$. $M_\mathrm{\rm Pl}=(G)^{-1/2}$ is the Planck
mass, $G$ is Newton's gravitational constant and for any quantity $X$
we denote time derivatives by $\dot{X}$. The condition for growth
$\ddot{a}>0\,,$ requires $p<-\rho/3$ and hence $\frac12
\dot{\phi}^2< V(\phi)$. CMB observations strongly favor the
slow-roll inflation $\frac12\dot{\phi}^2\ll V(\phi)$ condition and
hence the dark energy equation of state $w=p/\rho\simeq -1$. Indeed,
the Planck mission measured $w=-1.13^{+0.13}_{-0.10}\,$~\cite{Ade:2013zuv}. The
first Friedman equation reads
\bea
\dot{a}^2/a^2+k/a^2=\ell^2\,\rho\cs
\eea
and may be written as $H^2=\ell^2\,
\left[V(\phi)+\frac12\,\dot{\phi}^2\right]=\ell^2\,\rho$. The kinetic 
term $\dot{\phi}^2$ is controlled by $\dot{H}=-\frac32
\ell^2\,\dot{\phi}^2$ related to the observationally controlled
deceleration parameter $q(t)=-\ddot{a}a/\dot{a}^2$. In addition we have the field
equation 
\bea
\ddot{\phi}+3H\dot{\phi}=-V'(\phi)\equiv -\D V(\phi)/\D \phi\,.
\label{fieldeq}
\eea
It follows that the Higgs likely can be identified as the scalar field
which drives inflation provided $\ha\,\dot{\phi}^2\ll V(\phi)$. It is
precisely the quadratically enhanced mass term in the Higgs potential
which makes the Higgs a good inflaton candidate. A dominant mass term also
looks to imply the inflaton to represent essentially a free
field. This seems to be supported by recent Planck mission constraints
on non-Gaussianity~\cite{Ade:2013ydc}. 

\begin{figure}
\centering
\includegraphics[height=4.0cm]{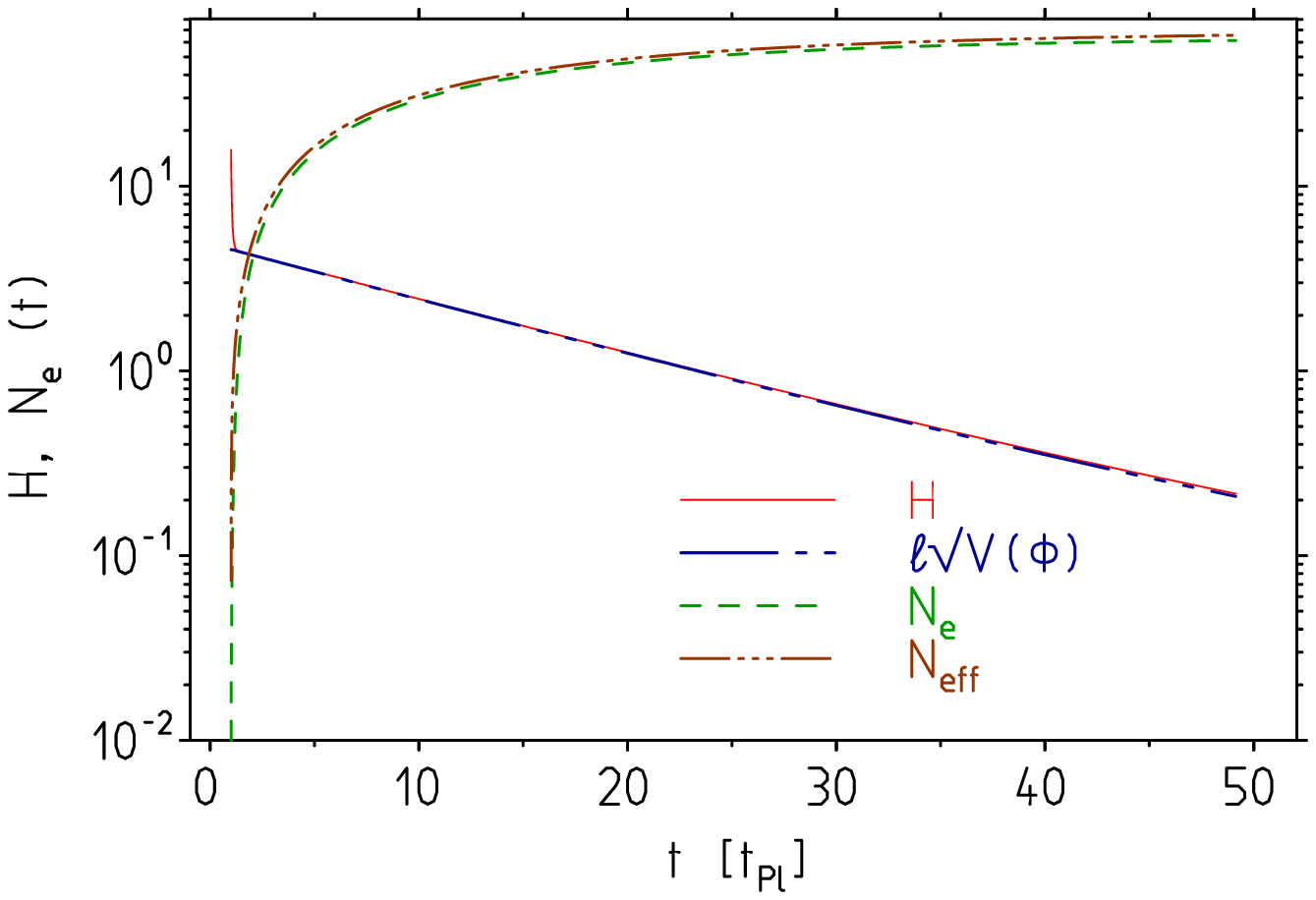}
\includegraphics[height=4.0cm]{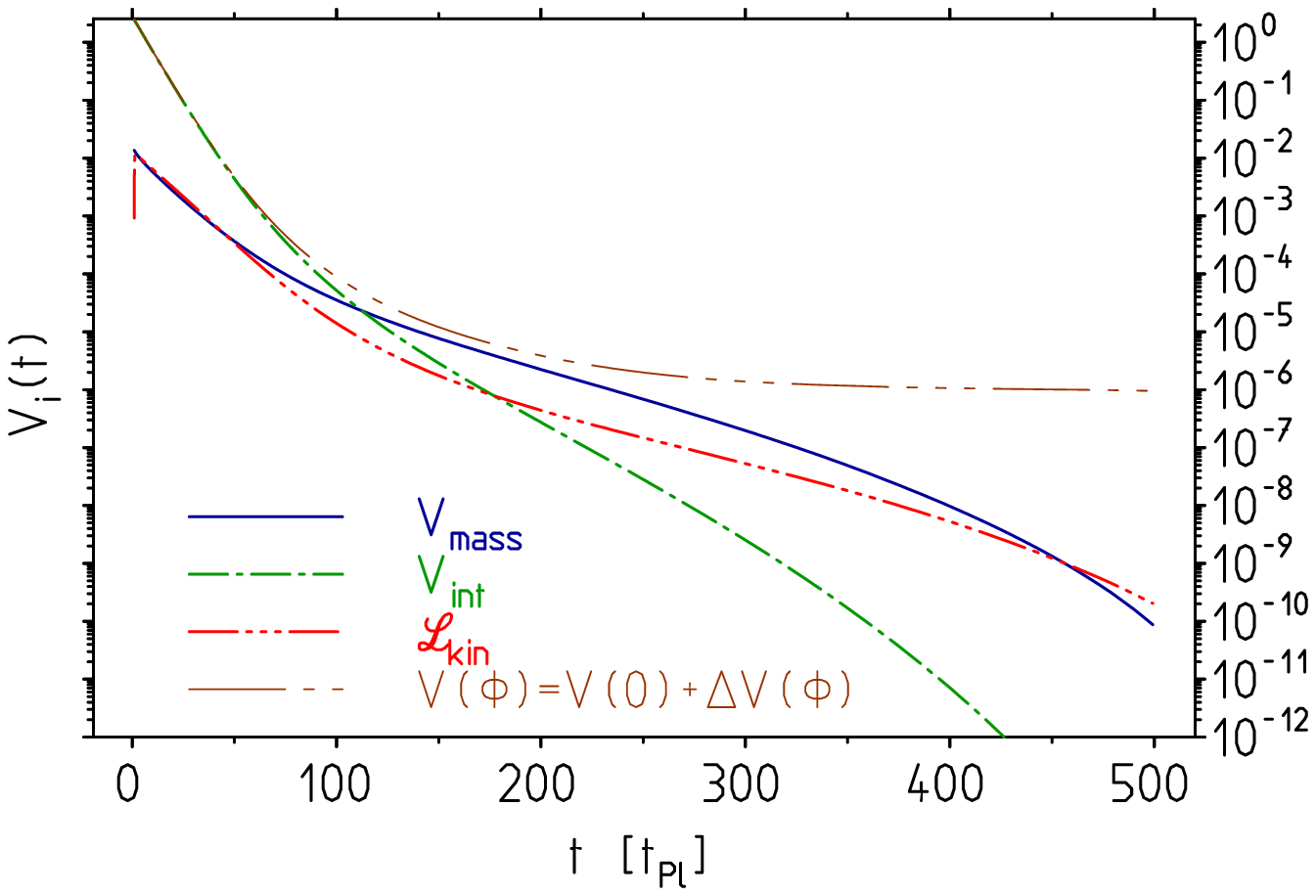}
\caption{The inflation era. Left: $N_e\approx 66$ is reached at time
$t\approx 50\,\tpl\,.$ The Hubble constant $H$ is satisfying
$H\approx \ell\,\sqrt{V(\phi)}$ very well shortly after Planck
time. The evaluation of $N_e$ via Eq.~(\ref{Neformula}) agrees very well with
the numerical $N_{\rm eff}=\ln a(t)/a(\tpl)$ obtained by solving the
coupled set of dynamical equations. Right: dark energy contributions
during inflation. Shown are the separate terms of the Higgs
potential together with the Higgs kinetic term. Slow-roll inflation
stops at about $t\approx 450\,\tpl$ when $\cL_{\rm kin} \sim V_{\rm
mass}\gg V_{\rm int}\,,$ after which the Higgs behaves as a damped
quasi-free oscillating field.}
\label{fig:HubbleInflation} 
\end{figure}

The amount of inflation is quantified by the inflation exponent $N_e$
given by
\bea
N_e &=&\ln \frac{a(t_{\rm end})}{a(t_{\rm
initial})}=\int\limits_{t_i}^{t_e}\,H(t)\,\D t
=\int\limits_{\phi_i}^{\phi_e}\,\frac{H}{\dot{\phi}}\,\D \phi \nn\\
&=&-\frac{8\pi}{\mpl^2}\int\limits_{\phi_i}^{\phi_e}\,\frac{V}{V'}\,\D
\phi =H\,(t_e-t_i)\,,
\label{Nedef}
\eea
where we have utilized the field equation $H \D t= -H^2/V'\,\D \phi$ and the first Friedman
equation $H^2=\ell^2 V$ in the slow-roll
approximation. The times $t_i$ and $t_e$ denote beginning and end of
inflation, where correspondingly the scalar field attains values $\phi_i$
and $\phi_e$, respectively.  For
\mbo{H=\mathrm{constant}} we would have
\mbo{N_e=H\,(t_e-t_i)}, which is a good approximation to the extent
that the total energy density \mbo{\rho_{\rm tot}\simeq\rho_\Lambda}
is dominated by the cosmological constant (CC). In the symmetric phase
\mbo{V/V'>0} and hence \mbo{\phi_i > \phi_e\,.}  A rescaling
of the potential does not affect inflation, but the relative weight of the
terms is crucial. A precise analysis of the relative importance of the
various possible components will be the main topic of the next Section.
For the SM Higgs
potential in the symmetric phase, denoting \mbo{z\equiv
\frac{\lambda}{6\,m^2}\,,} and a potential $V(\phi)=V(0)+\Delta V(\phi)$ we have a term 
\mbo{\frac{V(0)}{2m^2}\frac{1}{\phi}\frac{1}{1+z\phi^2}} plus
\mbo{\frac{\Delta V}{V'}=\frac{\phi}{4}\,\left(1+\frac{1}{1+z\phi^2}\right)}
and thus with
\ba
{\cal I}=\int\limits_{\phi_e}^{\phi_i}\,\frac{V}{V'}\,\D \phi=
\frac{V(0)}{2m^2}\left[\ln \frac{\phi_i^2}{\phi_e^2}-\ln \frac{
 {\phi_i}^2\,z+1}{{\phi_e}^2\,z+1}\right]+ 
\frac18\,\left[\phi_i^2-\phi_e^2
+\frac{1}{z}\,\ln \frac{
 {\phi_i}^2\,z+1}{{\phi_e}^2\,z+1}\right]
\ea
we obtain
\bea
N_e=\frac{8\pi}{\mpl^2}\,{\cal I}\epo
\label{Neformula}
\eea
Below we will show that
\bea
V(0)=\frac{m^2}{2}\, \langle 0 | \phi^2 |0\rangle + \frac{\lambda}{24}\,\langle 0 | \phi^4 |0\rangle
\eea
like $m^2$ and $z=\frac{\lambda}{6m^2}$ all are known SM quantities! 
\mbo{N_e} large requires \mbo{\phi_i \gg \phi_e}. With $\phi_i \simeq
4.51\, \mpl$, a value motivated by the amount of inflation
wanted, and taking into account the
\textbf{running} of parameters as given
by the standard \MSb RG, we find $\phi_e \simeq 2.01\power{-3} \mpl$
and $N_e \approx 64.68$ at the end of inflation at about $t\simeq
450\,\tpl\,,$ a value not far above the phenomenologically required
minimum bound. $N_e$ may be increased by increasing
$\phi_i$. Figure~\ref{fig:HubbleInflation} gives an overview already
of important features of the early inflation period.

As we will see, SM Higgs inflation is far from working obviously. The
reason why SM inflation is quite tricky is the fact that the form of
the potential is given and the parameters are known. What is at our
disposal is essentially only the value of the Higgs field at the
Planck scale, since in the experimentally accessible low energy region
the Higgs field is not an observable and we only know its vacuum
expectation value.

In the following we are dealing with physics near the Planck scale,
where the bare theory resides, and by $\phi,V(\phi),\lambda$ and $m$
we denote the bare quantities (fields and parameters), if not specified
otherwise.

The papers is organized as follows: in Section 2 we study in some
detail how Higgs inflation actually works. Section 3 is devoted to
considerations of reheating and the possibility of SM baryogenesis. A
solution of the cosmological constant problem of the SM is presented
in Section 4. Conclusions are following in Section 5.

\section{The profile of Higgs inflation}
In Ref.~\cite{Jegerlehner:2013cta} we have worked out the effective
structure of the SM up to the Planck scale, by calculating the bare SM
parameters in the \MSb scheme as a function of the energy scale
$\mu$. Above the Higgs transition scale $\mu_0$, in the early
universe, the SM is in the unbroken phase, where four heavy Higgs
fields exist besides the other highly relativistic degrees of
freedom. The analysis is strongly supporting that the Higgs is
actually responsible for the phenomenon of inflation.  Here we present
a more detailed investigation and perform consistency checks
concerning the SM inflation scenario (see
Ref.~\cite{Linde:1990xn} and references therein). So far we have not said much about the size
of the Higgs field and whether it is adequate to just check the
parameters in the potential to decide about the relative importance of
the different terms. Therefore some more details on the impact of the
quadratic enhancement on the inflation profile are in order. We have
shown that for a Higgs mass of about 126 GeV there is a phase
transition at a scale about $\mu_0
\sim 1.4 \power{16}~\gv$ and at temperatures above this scale the SM is
in the symmetric phase in which the Higgs potential exhibits a huge
bare mass term of size
\begin{equation}
m^2\sim \delta m^2 \simeq \frac{\mpl^2}{32\pi^2}\,C(\mu=\mpl) 
\simeq \left(0.0295\,\mpl \right)^2\,,\,\,
\end{equation}
or $m^2(\mpl)/\mpl^2\approx 0.87\power{-3}\epo$
For large slowly varying fields, the field equation of motion
simplifies to the slow-roll equation $3\,H\dot{\phi}\approx-V'$ with
$H\approx\ell \sqrt{V}$, which describes a decay of the field.

In the symmetric phase the key object of interest is the SM Higgs
potential
$V(\Phi)=m^2\,\Phi^+\Phi+\frac{\lambda}{3!}\,\left(\Phi^+\Phi\right)^2$
eventually dominated by the mass term
$m^2\,\Phi^+\Phi\,.$ Here $\Phi$ is the complex SM $SU(2)$ Higgs
doublet field, which in the symmetric phase includes four heavy
physical scalars:
\bea
\Phi=\left(\begin{tabular}{c}$\phi^+$\\
$\phi_0$\end{tabular}\right)\semis\phi^+=\I\,\frac{\phi_1-\I\,\phi_2}{\sqrt{2}}\comas
\phi_0=\frac{H-\I\,\phi}{\sqrt{2}}\comas\phi=\phi_3
\eea
in terms of the real fields $H,\phi_i\,,(i=1,2,3)$. In the broken
phase $\bra{0}{H}\ket{0}=v$, the $\phi_i$'s transmute to gauge degrees
of freedom and we get the Higgs potential
$V(H)=\frac{m^2}{2}\,H^2+\frac{\lambda}{24}\,H^4$, considered so
far. We adopt the ``would be'' charge assignments, as they manifest
themselves in the broken phase. In the symmetric phase $U(1)_{\rm em}$
is not yet singled out and there are no photons and in place of charge
and flavor there are the gauge symmetry assignments only, the
singlets, doublets and triplets.  Still hyper-charge $U(1)_Y$ is conserved.  We
will nevertheless use field assignments as if we would be in the
broken phase.

There are two quantities, which we cannot get by just extrapolating
the SM beyond the Higgs transition point. One is the
renormalized mass $m$ in the symmetric phase, the other is the
magnitude of the Higgs field. We assume here that the renormalized
$m^2$ is small relative to $\delta m^2$, which we can calculate. The
Higgs field in the LEESM only depends logarithmically on the cutoff,
such that we naturally expect the field the by small in the sense
$m^2\gg
\bra{0}\Phi^+\Phi\ket{0}$. In fact the Higgs fields must be very
heavy as well in order that the Higgs can be the inflaton. Here, the
field equation and the Friedman equations actually help to estimate
the proper initial value of the field. At least in the slow-roll
inflation scenario we know that the Higgs field has to decay fast. Obviously,
large Higgs fields work against Gaussianity, which requires the dominance
of the Higgs mass term, and hence
\bea
\phi^2 <
\frac{12\,m^2}{\lambda}=\frac{3\,C}{8\,\pi^2\,\lambda}\:\mpl^2=
\frac{3\,(2\lambda+\frac{3}{2}g'^2+\frac{9}{2}g^2-12y_t^2)}
{8\pi^2\lambda}\:\mpl^2
\eea
during inflation. With our input parameters, mass term dominance holds when
\bea
|\phi| \ll \sqrt{12/\lambda}\:m\:(\mpl) \approx 0.2726\,\mpl\epo
\label{phibound}
\eea
Since $\phi$ is decreasing rapidly during inflation, the condition of
Gaussianity gets dynamically established at some point, at which
however the dark energy density $V(\phi)$ has to be large enough to
keep inflation going.  We also note that RG evolution may yield a
substantially smaller value for $\lambda(\mpl)$ in case $y_t(M_t)$
would be slightly larger than our estimate (see
Table~\ref{tab:params}).  If $\lambda(\mpl)=0$, the minimum value
allowed for our scenario to work, we have $C(\mu) \leq 0$ for
$y_t(\mpl)>0.353$, and there is no Higgs transition below
$\mpl$. Thus, for our scenario to work we need $y_t(\mpl)<0.353$,
given the gauge couplings at $\mpl$.

In any case, it is very interesting that the whole scenario based on
the existence of the Higgs phase transition sufficiently below the
Planck scale, requires a window in parameter space which is very close
to whatever SM parameters estimates yield. Can this be an accident?
As our SM inflation scenario is supported by CMB data, we expect that
at the end something close to our scenario should turn out to describe
reality.

In the very early universe radiation is dominating the scene, this is
not changed by a large cosmological constant term, even the curvature
term may win over the cosmological constant close enough to the Planck
time. The Hubble constant in our scenario, in the symmetric phase,
during the radiation dominated era is given by $H=\ell
\sqrt{\rho}\simeq 1.66\,\left(k_B T\right)^2\sqrt{102.75}\,\mpl^{-1}$
such that at Planck time $H_i \simeq 16.83\,\mpl\epo$ One expects that
$V(\phi)$ does not exceed too much a possible vacuum energy of size $\mpl^4\,.$
The condition $V(\phi_0)=\mpl^4$ yields an initial value $\phi_0$ as
follows: with $a=6m^2/\lambda$, $b=24 \mpl^4/\lambda$ and
$r=\sqrt{b+a^2}$ we find 
\bea
\phi_0=\sqrt{r-a}\simeq 4.40\power{19}~\gv =3.61 \mpl \epo
\label{phinull}
\eea
It is well known that the CMB horizon problem requires an inflation index
$N_e > 60$. This index may be considered as a direct measure of the unknown initial
value $\phi_0$, and hence observational inflation data actually
provides a lower bound on this input. With the plausible estimate (\ref{phinull})
we actually obtain $N_e \sim 57$ but we easily can reach an index
above $N_e\sim 60$ by slightly increasing (\ref{phinull}). We will
adopt an initial field enhanced by 25\% i.e. as our standard input we choose 
\bea
\phi_0=4.51 \mpl\epo
\label{phinullplus}
\eea
Notice that the dominance of the mass term at $\mpl$ would require 
the condition (\ref{phibound}).
The assumption here is that the Higgs field, having dimension one, in
the cutoff system with intrinsic scale $\mpl$ should
naturally be $O(\mpl)$. In fact we assume that the Friedman
equations as well as the Higgs field equations start to be valid after
$\tpl$ only, which does not mean that at earlier times temperatures $T
> T_{\rm Pl}$ and corresponding excitations of the system are not expected
to exist. They exist in any case. As the field decays exponentially at first, when
$V_{\rm int} \gg V_{\rm mass}$, during early 
inflation (see Fig.~\ref{fig:HubbleInflation} and Fig.~\ref{fig:Lagrangian} below) the mass term dominance (free massive field as an
inflaton) is reached in any case during inflation at times $t\gapprox 150\,\tpl\,.$ Once $V_{\rm
mass} \gg V_{\rm int}$, the field continues to decay exponentially, because
of the vacuum term $V(0)$, which largely determines the Hubble rate $H$. 

\begin{figure}
\centering
\includegraphics[height=4cm]{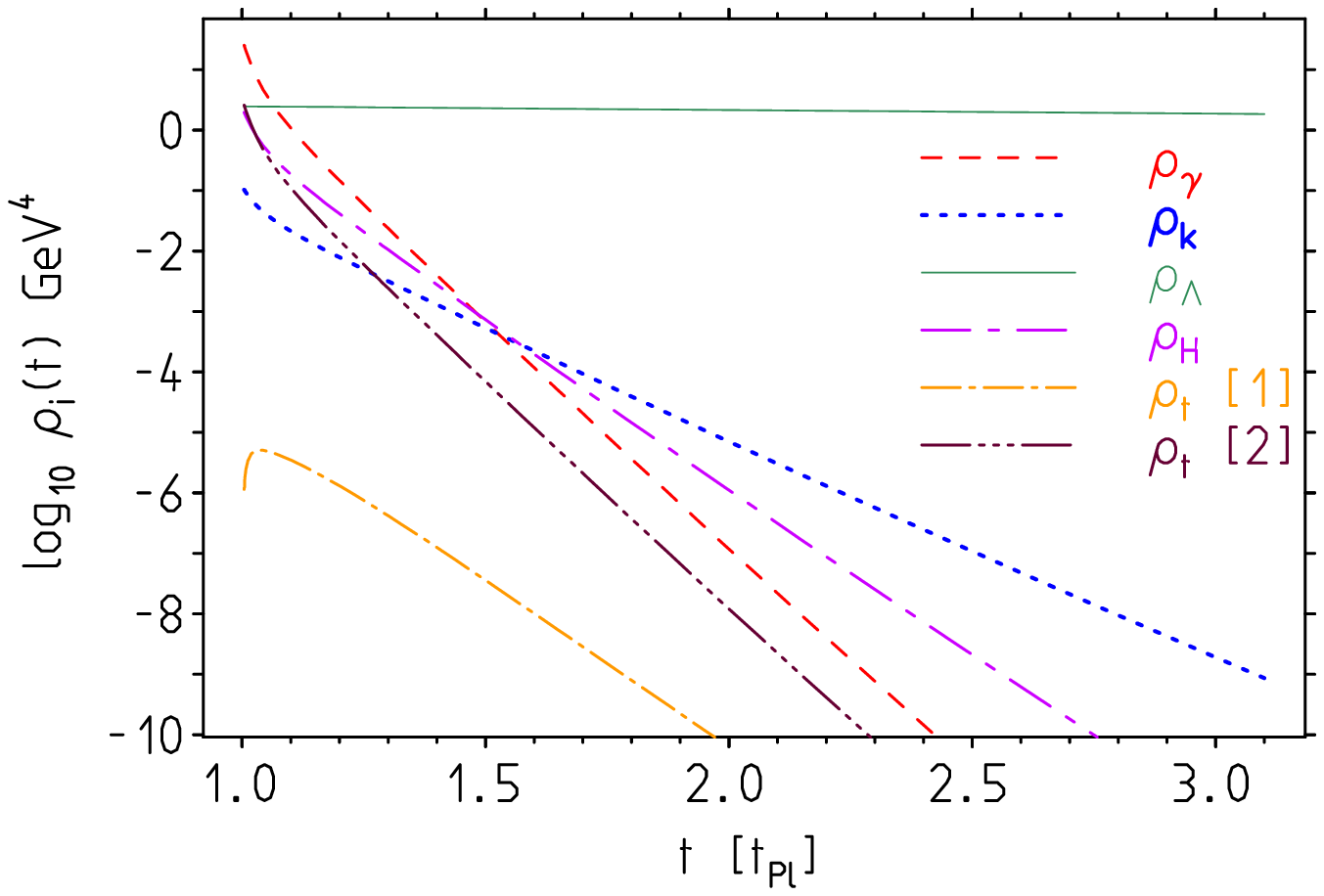}
\includegraphics[height=4cm]{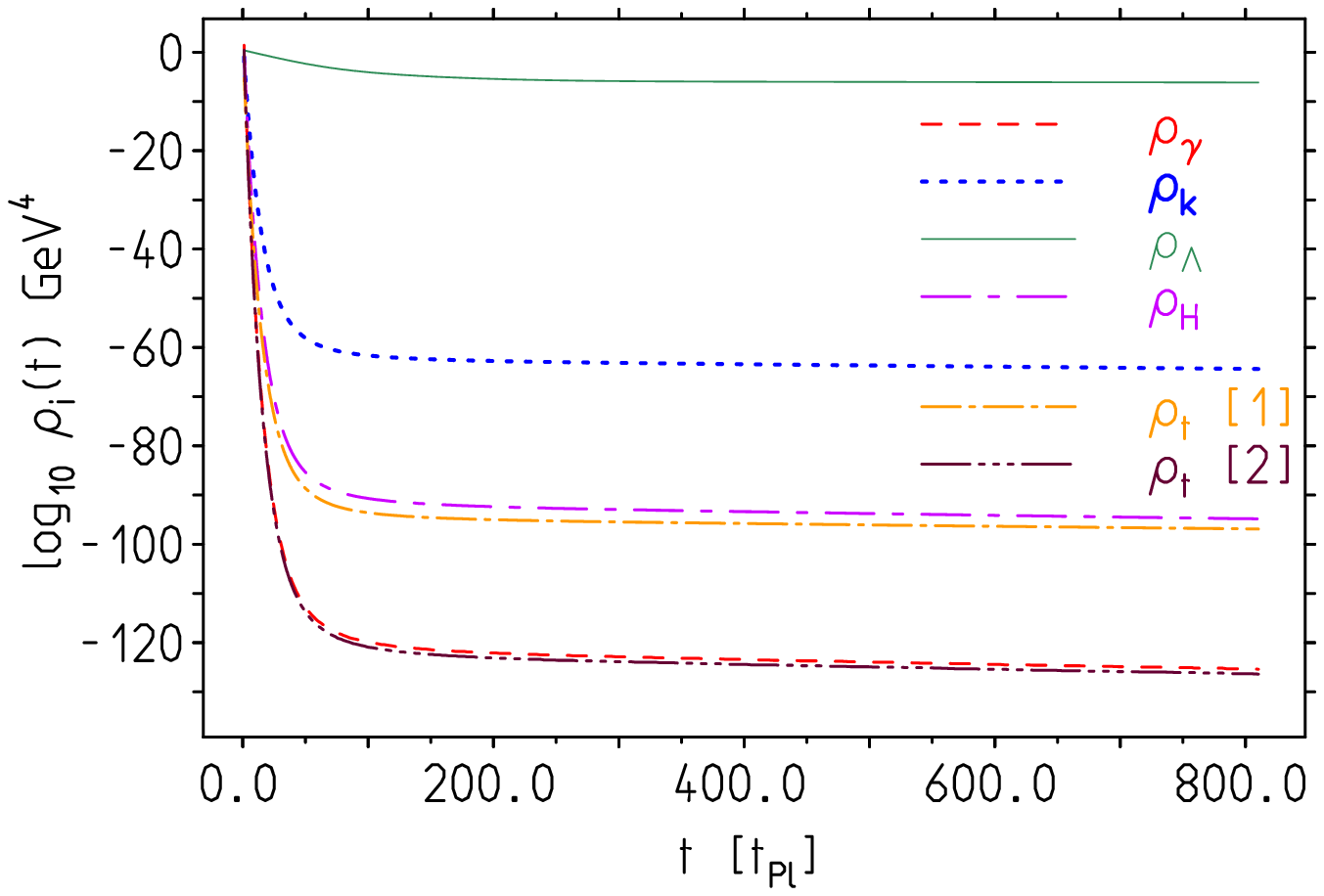}
\caption{Right: this graph shows that the dark energy term as given by
(\ref{barem2}) is dominating the density quite soon after the big
bang, and thus indeed is driving inflation. Left: analogous plot
taking into account the Higgs decay into $t\bar{t}$ [1] up to before the Higgs transition
($\phi_0\simeq 4.5\,\mpl$). $\rho_t$ [2] (in the plot little below
$\rho_\gamma$) is the top component in
$\rho_\gamma$, which denote the total radiation density without the
reheating part. For later reference we also show the top quark
density $\rho_t$ obtained from reheating [1] in comparison with the
fraction of top quark radiation which is part of
$\rho_\gamma$. Included is also the non-relativistic heavy Higgs
component $\rho_H$ given by Eq.~(\ref{HeavyHiggsMatter}).}
\label{fig:rhooft} 
\end{figure}

In order to get an idea on how the different density components affect
the evolution at very early times we follow
Refs.~\cite{Harrison:1967zz,GalindoDellavalle:2008dg}.  Let us rewrite
the 1$^{\rm st}$ and 2$^{\rm nd}$ Friedman equations in the form
\begin{eqnarray}
\left(\frac{\dot{a}}{a}\right)^2 = \ell^2\,\left(\rho_\gamma+\rho_\Lambda-\rho_k\right)
~~\mathrm{ \ and \ }~~
\frac{\ddot{a}}{a}  =  \ell^2\,\left(-\rho_\gamma+\rho_\Lambda\right)\cs
\label{fried12}
\end{eqnarray}
where
\begin{eqnarray}
\rho_\gamma=\rho_{\gamma i}\,\left(\frac{a_{i}}{a}\right)^4\semis
\rho_\Lambda=\frac{\Lambda}{3\ell^2}\semis
\rho_k=\frac{k c^2}{\ell^2 a^2}\cs
\label{therhos}
\end{eqnarray}
represent the radiation, cosmological constant $\Lambda$ and curvature
contribution to the energy densities, respectively. Indexed by $i$ are the
corresponding initial quantities at the Planck scale. Matter will be produced
by reheating and at the EW phase transition at a later stage. By
adding the two Friedman equations one obtains
\ba
\frac12\,\frac{\D^2}{\D t^2}\,a^2(t)=2\ell^2\,\rho_\Lambda
\,a^2-k c^2\,,
\ea
which may be written as
\ba
\ddot{X}=E_\Lambda^2\,X\semis X\equiv a^2-2 k c^2 t^2_\Lambda 
\semis E_\Lambda=2\ell\sqrt{\rho_\Lambda}=1/t_\Lambda\,,
\ea
with solution
\ba
a(t)=\left(c_1\,\E^{-t E_\Lambda}+\frac12\,c_2\,\E^{t E_\Lambda}+2 k
c^2\,t^2_\Lambda \right)^{1/2}\epo
\ea
The integration constants $c_1$ and $c_2$ may be fixed by assuming
special initial values for $a_i=a(t_i)$ and $H_i=\dot{a}_i/a_i$ at the initial
time $t_i$, which we choose to be the Planck time. One obtains
\ba
c_1=\frac12\, \left(a_i^2-2\dot{a}_ia_i t_\Lambda
-2kc^2t_\Lambda^2\right)\,\E^{t_i E_\Lambda}\semis
c_2=\left(a_i^2+2\dot{a}_ia_i t_\Lambda
-2kc^2t_\Lambda^2\right)\,\E^{-t_iE_\Lambda}\,,\nn 
\ea
where $\dot{a}_i$ can be calculated in terms of the initial densities
$\rho_{\gamma i}$, $\rho_{k i}$ and $\rho_{\Lambda}$, which yields
\ba
\left(2 t_\Lambda \dot{a} a\right)_i=a_i^2\,\left(\frac{\rho_{\gamma
i}}{\rho_\Lambda}-\frac{\rho_{ki}}{\rho_\Lambda}+1 \right)^{1/2}\semis
2kc^2t^2_\Lambda=\frac12 a_i^2\, \frac{\rho_{ki}}{\rho_\Lambda}\epo
\ea
The solution then reads
\bea
a(t)=a_i\,\left(\cosh \tau + \frac12\,\frac{\rho_{ki}}{\rho_\Lambda}
\,\left(1-\cosh \tau \right) +
\left(\frac{\rho_{\gamma
i}}{\rho_\Lambda}-\frac{\rho_{ki}}{\rho_\Lambda}+1\right)^{1/2}\,\sinh \tau\right)^{1/2}
\eea
where $\tau=(t-t_i)\,E_\Lambda$ is the reduced time. The Hubble
function is then
\bea
H(t)=\frac{E_\Lambda}{2}\left(\frac{a_i}{a}\right)^2\,\left(\left(1-\frac12
\frac{\rho_{ki}}{\rho_\Lambda}\right)\,\sinh \tau +\left(\frac{\rho_{\gamma
i}}{\rho_\Lambda}-\frac{\rho_{ki}}{\rho_\Lambda}+1\right)^{1/2}\,\cosh \tau\right)\,,
\eea
while the acceleration reads
\bea
\frac{\ddot{a}}{a}=\frac{E_\Lambda^2}{2}\,\left(1-\frac12 \frac{\rho_{ki}}{\rho_\Lambda}\left(\frac{a_i}{a}\right)^2\right)-H^2\epo
\eea
We can then calculate the evolution of the various components which we
show in Fig.~\ref{fig:rhooft} together with the top quark density and the
heavy Higgs density to be introduced later.

As an initial value for the Higgs field at $\mpl$ we adopt
(\ref{phinullplus}).  In general, if the initial value of \mbo{\phi}
is exceeding about \mbo{\frac15 \,\mpl} one can neglect
\mbo{\ddot{\phi}} in the field equation as well as the kinetic term
\mbo{\frac12\,\dot{\phi}^2} in the Friedman equation. We expect
inflation to start at Planck time
\mbo{t_i\equiv t_\mathrm{initial}=\tpl\simeq 5.4 \power{-44}~\mathrm{sec}} 
and to stop definitely at $t_{\rm CC}\simeq 2.1
\power{-40}~\mathrm{sec}$ the at drop of the CC  to be discussed in
Sec.~\ref{Sect:CC}. As mentioned earlier the efficient era of slow-roll
inflation ends at about $t\simeq 450\,\tpl$. We adopt, somewhat
arbitrary, a period including the bare  Higgs transition point
at \mbo{t_e\equiv t_{\rm end}=t_{\rm Higgs}\approx 4.7
\power{-41}~\mathrm{sec}\,.} 
As $m$ is substantially lower than $\mpl$ actually for strong fields
the interaction term is dominating. Then $\phi$ decays exponentially like
\bea
\phi(t)=\phi_0\,\E^{-E_0\,(t-t_0)}\,;\, 
E_0=\frac{\sqrt{2\lambda}}{3\sqrt{3}\ell}\approx4.3\power{17}~\gv;\,
V_{\rm int} \gg V_{\rm mass}\,,
\label{phidecay}
\eea 
while a dominant mass term leads to a decay linear in time
\bea
\phi(t)=\phi_0-X_0\,(t-t_0)\,;\,
X_0=\frac{\sqrt{2}m}{3\ell}\;\approx7.2\power{35}~\gv^2;\,
V_{\rm mass} \gg V_{\rm int}\,.
\eea
However, this is not quite what corresponds to the true SM prediction.
As we will argue below we have to account for a cosmological constant term
$V(0)\equiv\braket{V(\phi)}$, which in the field equation contributes
the Hubble constant as $H\simeq\ell\,\sqrt{V(0)+\Delta
V(\phi)}$. At the begin of inflation $V(0)$ is of size comparable to
$V_{\rm int}$ while later on the mass term starts to dominate over the
interaction term, but still 
$V(0)\gg V_{\rm mass}$, such that actually also during this era we have
an exponential decay 
\bea
\phi(t)\approx \phi_0\,\E^{-E_0\,(t-t_0)}\,;\, 
E_0\approx\frac{m^2}{3\ell\,\sqrt{V(0)}}\approx6.6\power{17}~\gv;\,
V_{\rm mass} \gg V_{\rm int}
\label{phidecay2}
\eea 
of the field.
Thus in any case, during slow-roll inflation, the decay of the
dynamical part of the Higgs field is exponential at dramatic
rate. Actually, as we will see, the cosmological constant proportional
to $\rho_\Lambda = V(\phi)\approx V(0)$ and a corresponding Hubble
constant $H\approx\ell
\sqrt{V(0)}\,$ long after slow-roll inflation has ended, will decrease
dramatically when $V(0)$ drops essentially to zero at a scale $\mu_{\rm
CC}\simeq 5.0\power{15}~\gv\,.$ Without the contribution $V(0)$, the
fast decay of the Higgs field could be in contradiction with the
observationally favored slow-roll scenario. An important point here is
that the Higgs potential has a calculable non-vanishing vacuum
expectation value in the bare system.  Vacuum contractions are ruled
by Wick ordering\footnote{Wick ordering amounts to a redefinition of
the operator basis by subtracting c-number self contractions of the
fields. In gauge theories one is advised not to express the Lagrangian
in terms of Wick ordered fields because the originally manifest
symmetry would be mixed up and become intransparent. Wick ordering in
general is not just equivalent to subtracting the VEV of the
Lagrangian. }: $\phi^2=\braket{\phi^2}+{\phi'}^{2}$ and
$\phi^4=\braket{\phi^4}+3!\,\braket{\phi^2}{\phi'}^{2}+{\phi'}^{4}$
which yields a constant $V(0)=\braket{V(\phi)}$ and a mass shift
${m'}^{2}=m^2+\frac{\lambda}{2}\braket{\phi^2}$ plus the potential in
terms of the fluctuation field $\Delta V(\phi)\,.$ Note that in SM notation
$\bra{0}\Phi^+\Phi\ket{0}=\frac12\bra{0}H^2\ket{0}\equiv
\frac12\,\Xi$ is a singlet contribution.
We thus obtain a quasi-constant vacuum density
\bea
V(0)=\frac{m^2}{2}\,\Xi+\frac{\lambda}{8}\,\Xi^2\semis
\Xi=\frac{\mpl^2}{16\pi^2}\,,
\label{theCC}
\eea 
the VEV of the potential, which does contribute to the cosmological
constant.  The field equation, which only involves the time dependent
part, is affected via a modyfied Hubble constant and the shifted
effective mass (see below). The $Z_2$ symmetry $\Phi \to -\Phi$ and
the SM gauge symmetry remain untouched. The $\Phi^+\Phi$ VEV
$\frac12\,\Xi$ in principle should be calculable in a lattice SM. One
has to be aware of course that we do not know the true underling
Planck ether system. Such estimates in any case would be instructive
in understanding underlying mechanisms. Since, as a result of the SM
RG evolution, effective SM parameters are well within the perturbative
regime one can actually calculate $\Xi\,$.  In leading order we just
have Higgs self-loops\\
\centerline{\includegraphics[height=10mm]{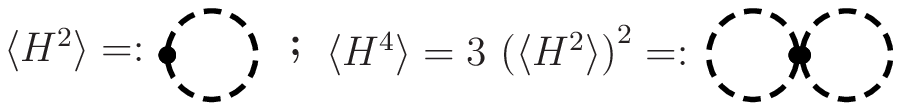}}
given by
$\Xi=\braket{H^2}=\frac{\Lambda^2}{16\pi^2}\,$\footnote{Including the
heavy Higgs mass effect we have
$\Xi=\frac{\mpl^2}{16\pi^2}\,\left(1-\frac{m^2}{\mpl^2}\,\ln
\left(\frac{\mpl^2}{m^2}+1\right)\right)$. This mass correction is
negligible here.}. Again, for the fluctuation field, which decays
exponentially, in the early phase of inflation we adopt the initial
value $\phi_0\approx 4.51 \mpl$ estimated above in
Eq.~(\ref{phinullplus}). In the potential of the fluctuation field and
the corresponding field equation the mass square now is given by
\bea
{m'}^{2}=m^2+\frac{\lambda}{2}\,\Xi\epo
\label{massshift}
\eea
This actually is a very interesting shift as it modifies the Higgs
transition point to lower values with new effective coefficient
\bea
C'_1=C_1+\lambda=3\,\lambda+\frac32\, {g'}^{2}+\frac92\,g^2-12\,y_t^2\,.
\label{coefC1prime}
\eea 
For our values of the \MSb input parameters, we
obtains (see Fig.~\ref{fig:lambdashiftFT} below)
\bea
\mu_0 \approx 1.4 \power{16}~\gv \to \mu'_0\approx 7.7 \power {14}~\gv\,,
\label{newPTP}
\eea 
as a relocation of the Higgs transition point.

\begin{figure}
\centering
\includegraphics[height=4cm]{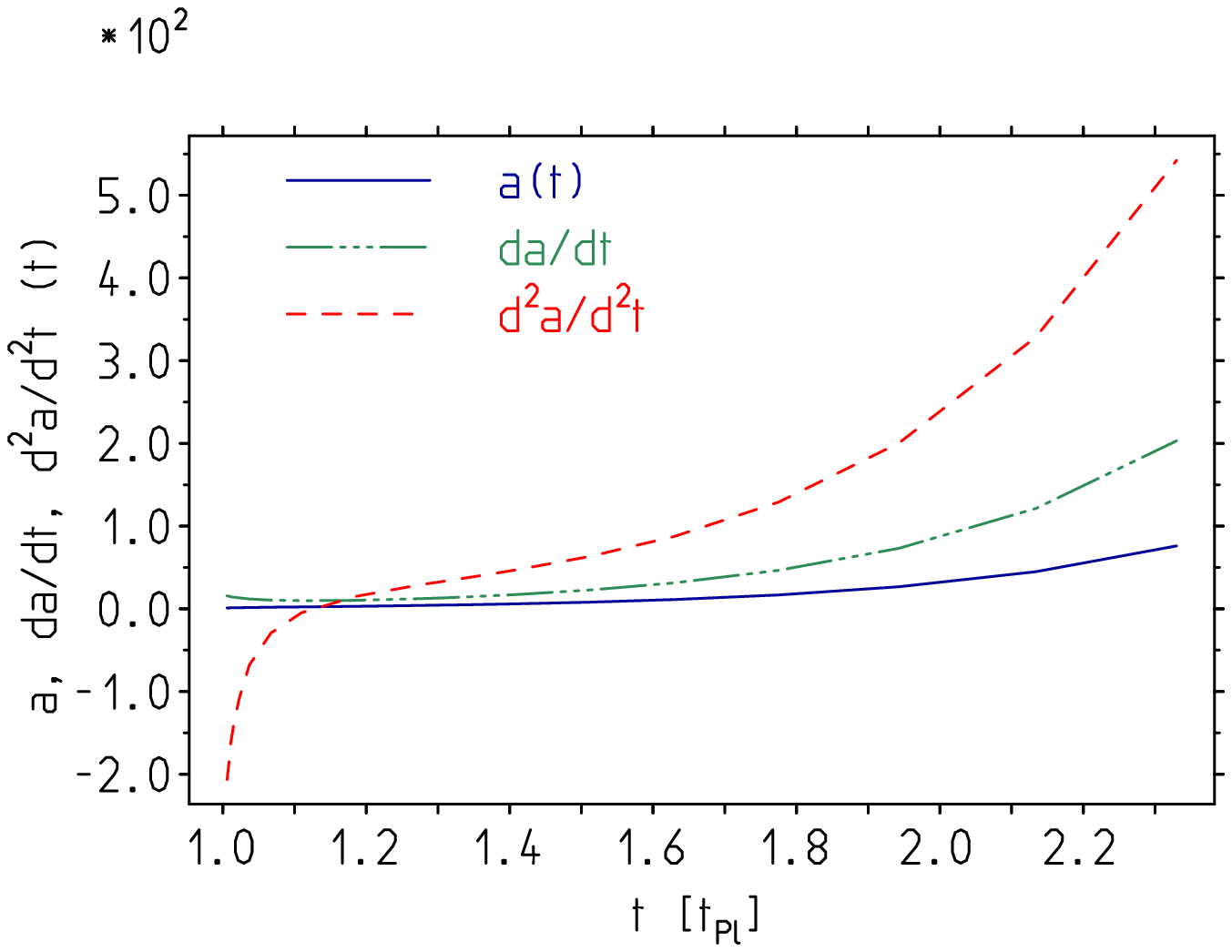}
\includegraphics[height=4cm]{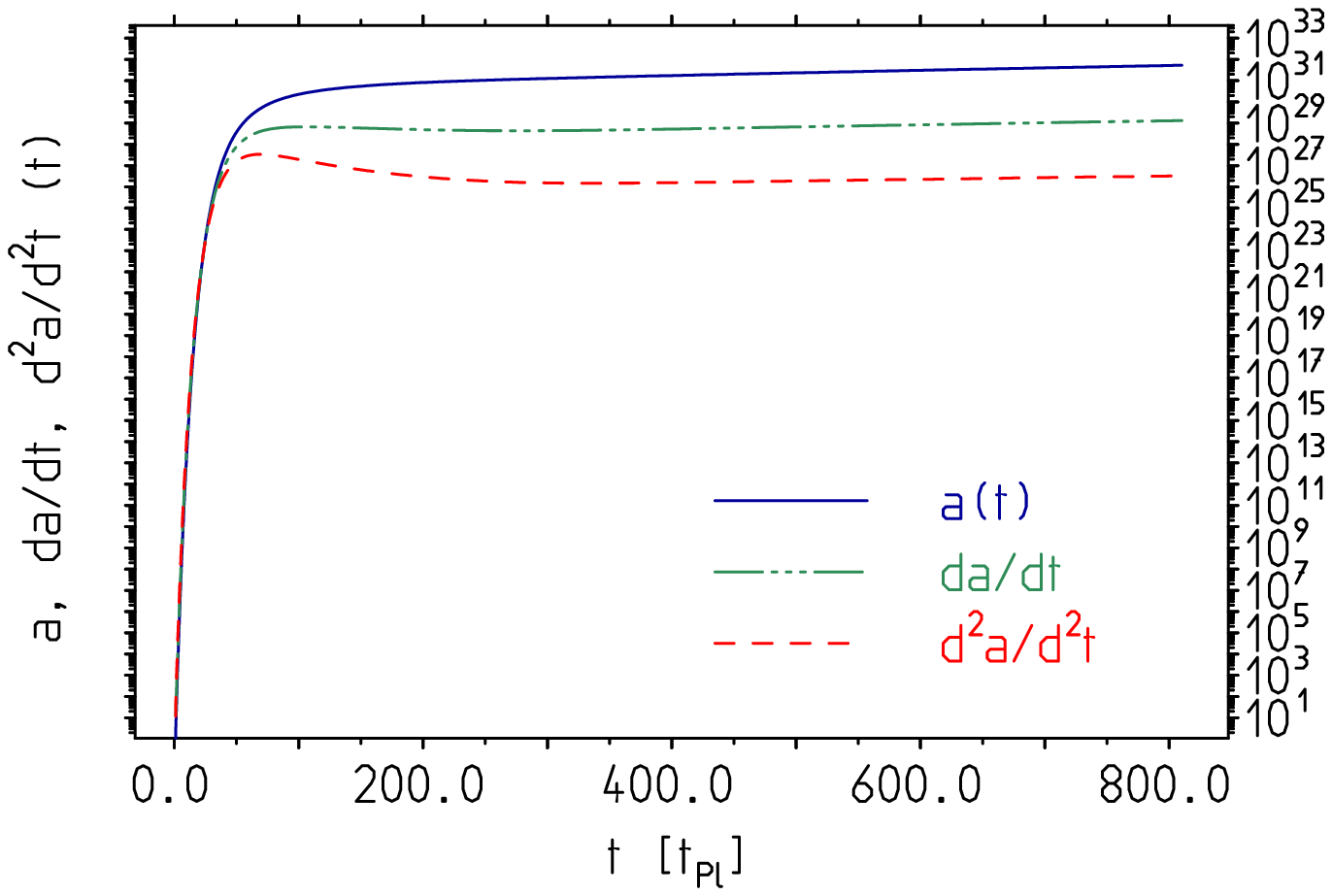}
\caption{The FRW radius and its derivatives for $k=1$ as a function of
time all in units of the Planck mass, i.e. for $\mpl=1$. Left: the
start. Right: the expansion before the Higgs transition.}
\label{fig:aoft} 
\end{figure}

\begin{figure}
\centering
\includegraphics[height=4cm]{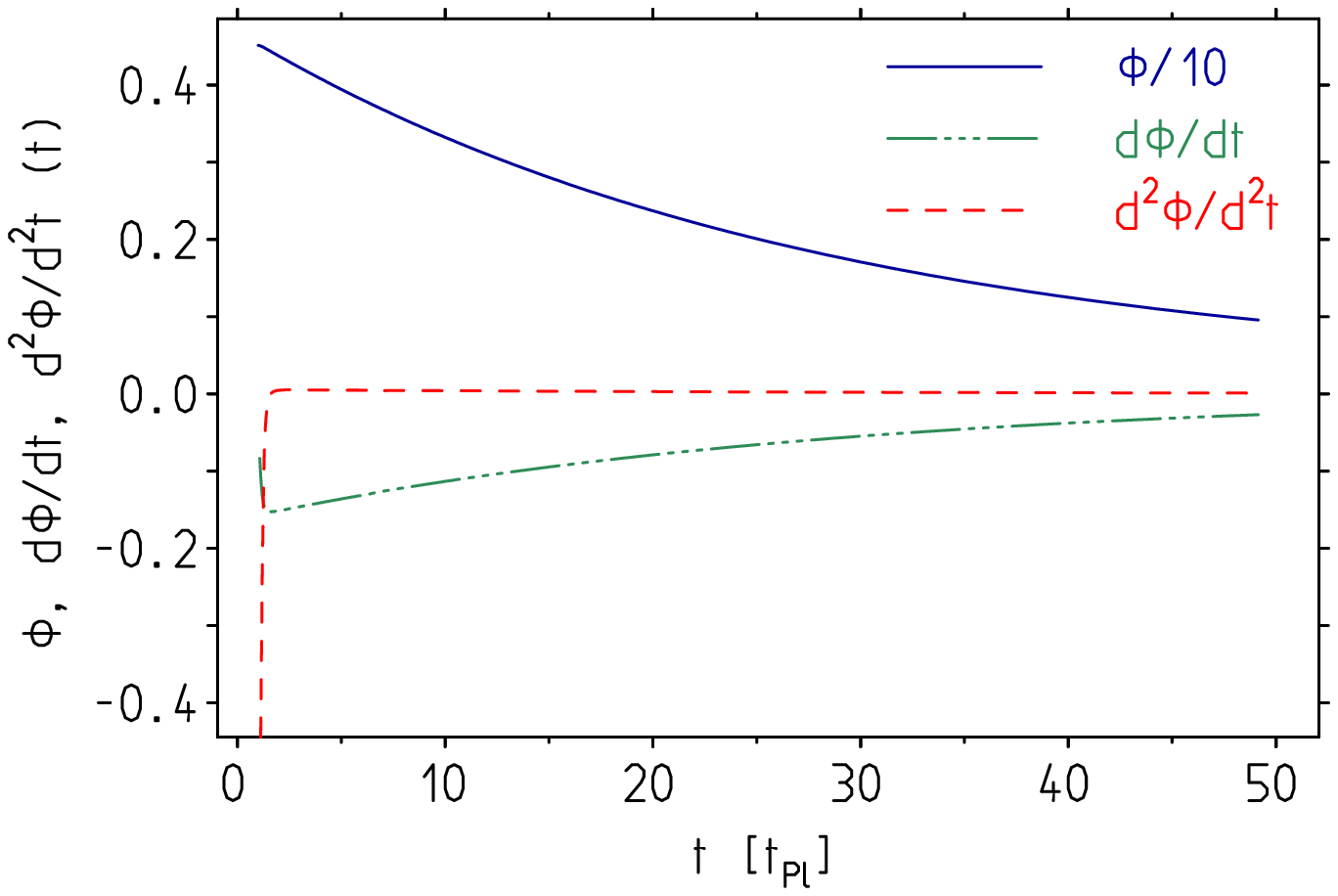}
\includegraphics[height=4cm]{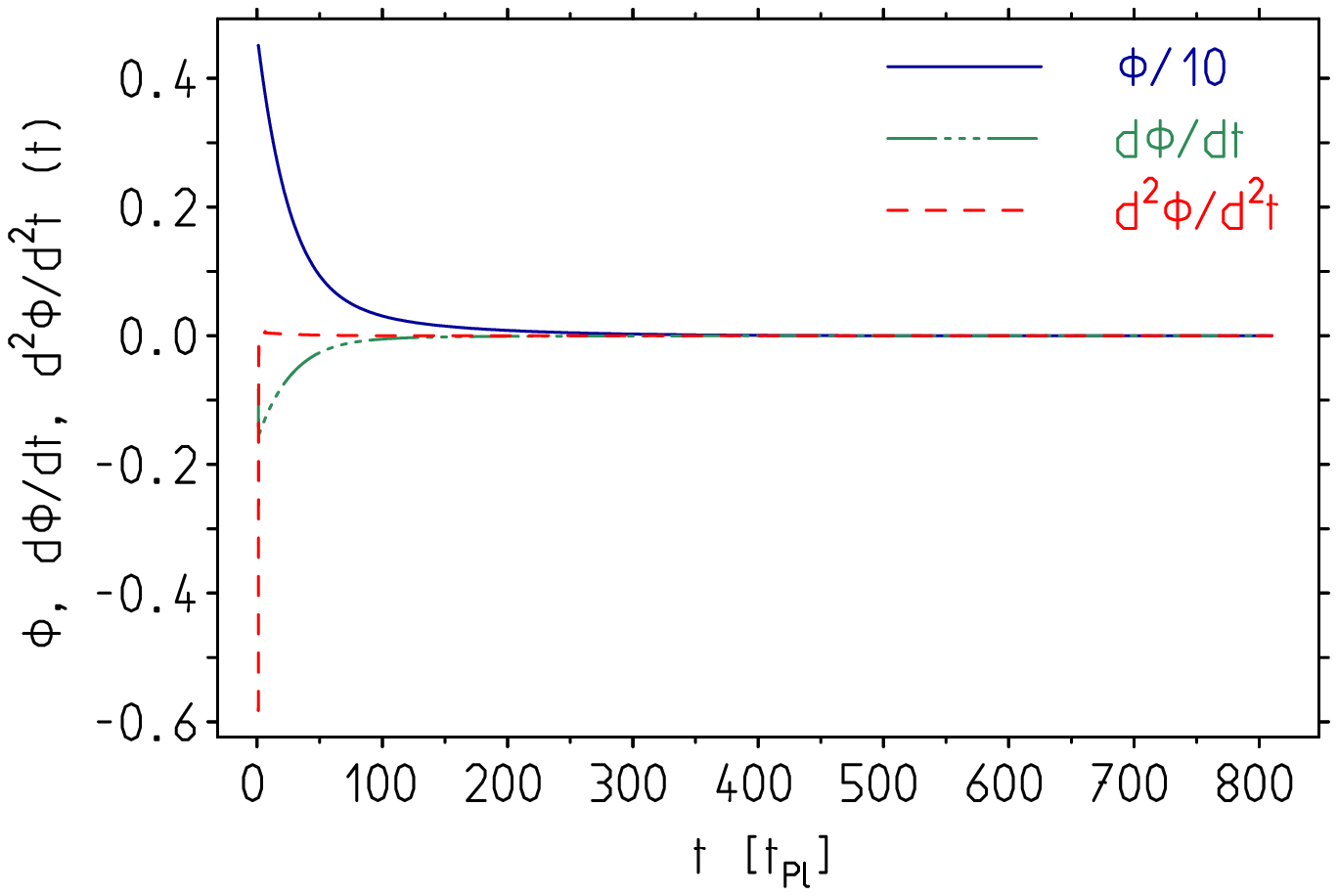}
\caption{The Higgs field and its derivatives for $k=1$ as a function of
time all in units of the Planck mass. Left: the Higgs
field at start. Right: the Higgs field decay before the Higgs transition.
The field starts oscillating strongly like a free field once $\cL_{\rm
kin}\sim V_{\rm mass}$ while $V_{\rm int} \ll V_{\rm mass}$ (see
Fig.~\ref{fig:Lagrangian}).}
\label{fig:field} 
\end{figure}

We can now solve the coupled system of equations
(\ref{fried12},\ref{therhos}) and (\ref{fieldeq}) numerically e.g. by
the Runge-Kutta method. It adds to the above analytic solution for a
constant ``cosmological constant'' the Higgs field dynamics. Results
for the FRW radius $a(t)$ and the field $\phi(t)$ together with the
derivatives are displayed in Figs.~\ref{fig:aoft},\ref{fig:field}.
\begin{figure}
\centering
\includegraphics[height=4cm]{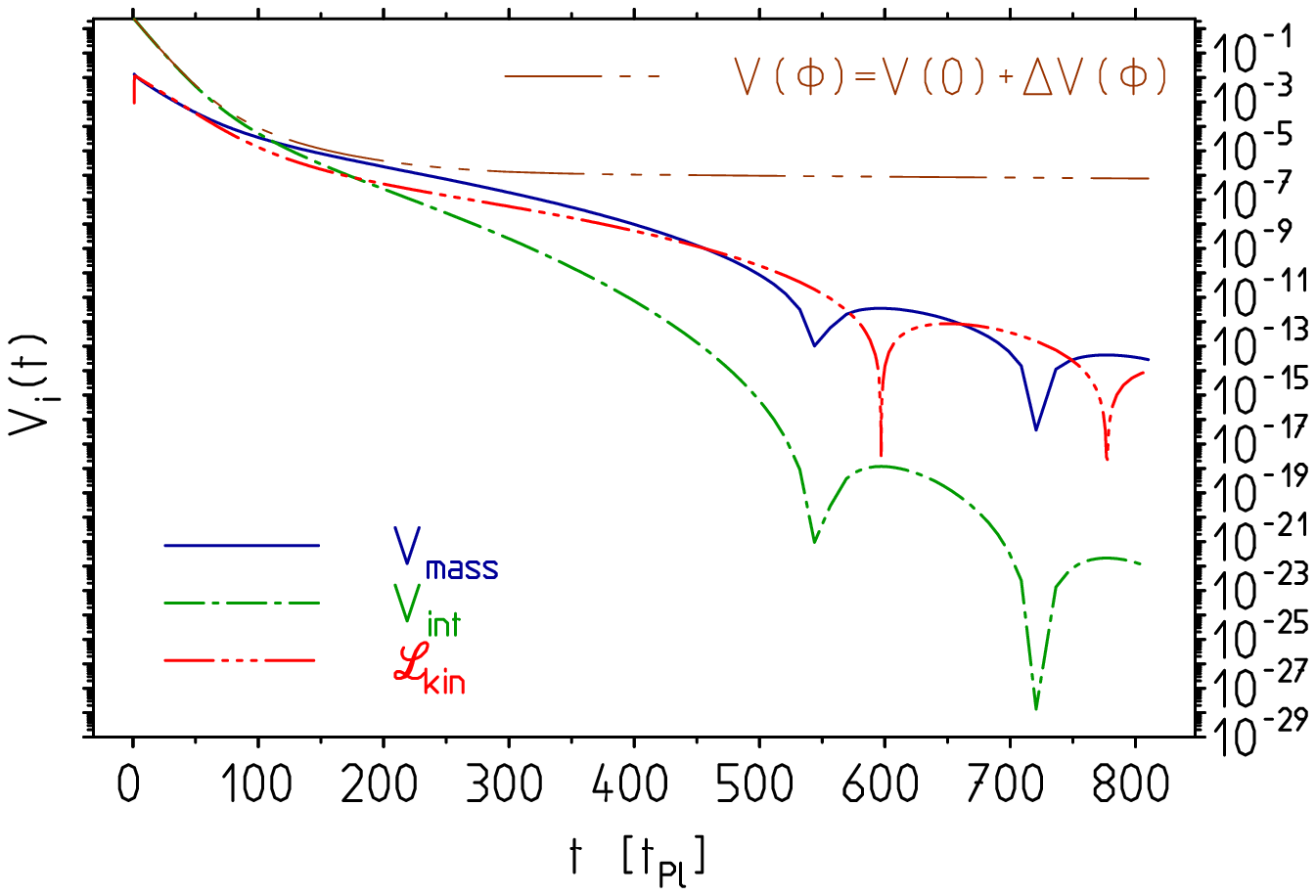}
\includegraphics[height=4cm]{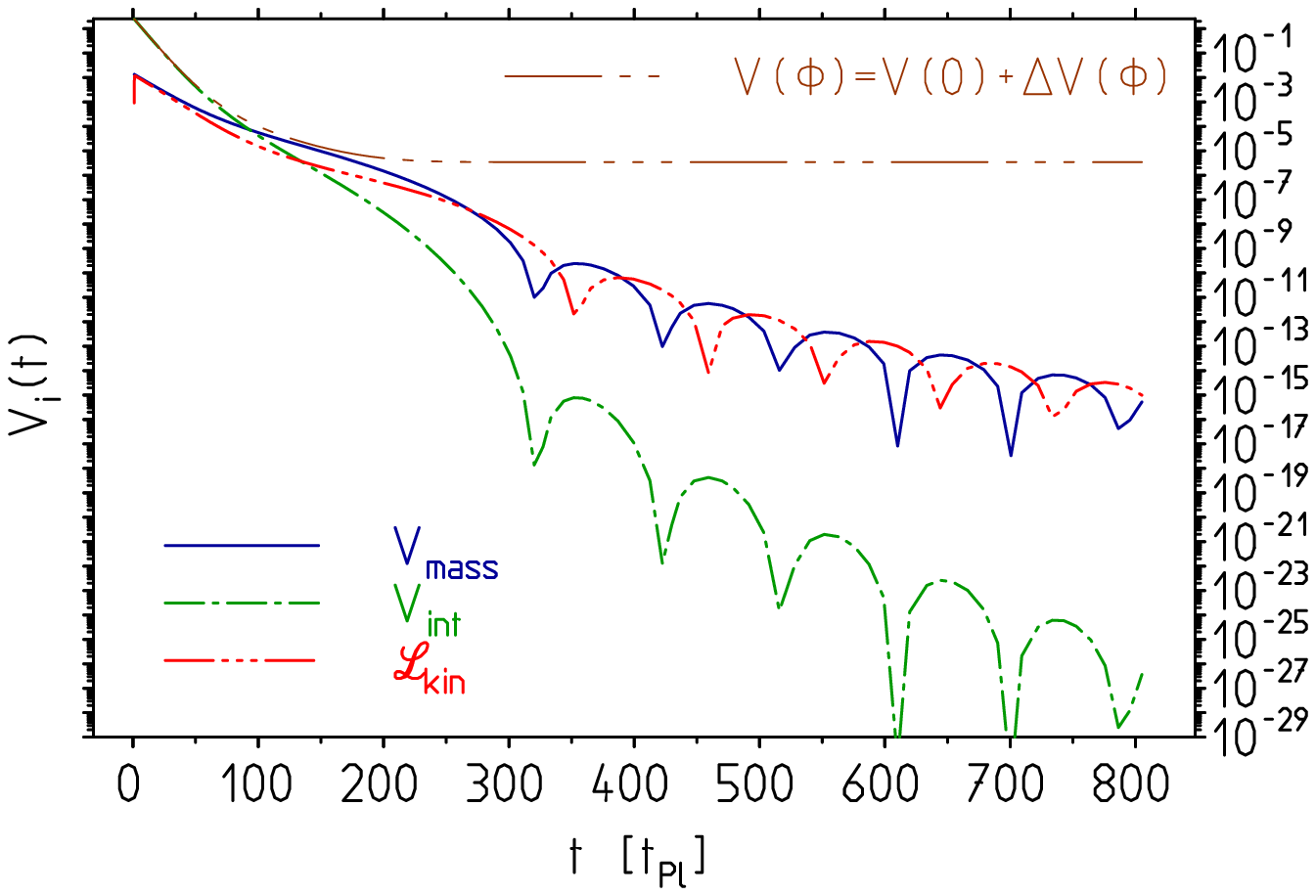}
\caption{The mass-, interaction- and kinetic-term of the bare
Lagrangian in units of $\mpl^4$ as a function of time. Left: the
relative contributions when proper running of SM couplings are taken
into account. The mass term is dominating in the range $t\simeq 100
\mathrm{ \ to \ } 450\,\tpl\,,$ where the slow-roll era ends and
damped quasi-free field oscillations start. Right: fake solution based on
constant couplings.}
\label{fig:Lagrangian} 
\end{figure}
Figure~\ref{fig:Lagrangian} shows how the different terms of the bare
Lagrangian evolve. At later inflation times the mass term is
dominating as originally expected, but the dominance is not very
pronounced. The temporary mass term dominance is important for the
observed Gaussianity by the Planck mission~\cite{PlanckResults}.  At
about $t\simeq 450\,\tpl\,$ slow-roll inflation ends and free field
oscillations begin. The two panels illustrate the difference obtained
between working with running couplings vs. keeping couplings fixed as
given at the Planck scale. It turns out to be crucial to take into
account the scale dependence of the coupling, throughout the
calculation. How can it be that the minor changes in SM couplings
between $\mpl$ and $\mu_0$ make up such dramatic difference?  The
reason is quite simple, the effects are enhanced by the quadratic
``divergence'' enhancement factor $\frac{\mpl^2}{32\pi^2}$, which
actually makes the whole thing work.  One of the key criteria during
the inflation era is the validity of the dark energy equation of state
$w=p/\rho=-1$, which we display in Fig.~\ref{fig:eqofstate} as a
function of time before the bare Higgs transition point $\mu_0$. In fact
$w=-1$ is perfectly satisfied quite early after Planck time.  This
shows how dark energy is supplied by the Higgs system.

The slow-roll criteria are usually tested by the coefficients
\bea
\eps \equiv
\frac{\mpl^2}{8\pi}\,\frac12\,\left(\frac{V'}{V}\right)^2\semis
\eta \equiv \frac{\mpl^2}{8\pi}\,\frac{V^{''}}{V}\,
\label{indicesdef}
\eea
where $\eps \ll 1$
ensures $p \simeq -\rho $, while \mbo{\eps,\eta \ll 1} ensure
slow-roll for a long enough time, maintaining
\mbo{\ddot{\phi}\ll3\,H\dot{\phi}\,.}  When slow-roll ends, \mbo{\phi}
oscillates rapidly about \mbo{\phi=0} and the oscillations lead to
abundant particle production which is reheating the universe. More on
this below.

Actually, as we will see, the condition $\eta \ll 1$ is a sufficient
condition only and not a necessary one.  For a SM Higgs type potential
especially $\eta \ll 1$ is hard to satisfy as we will explain
below. In the LEESM scenario $V(0)$ plays a crucial role and actually
keeps inflation going on in spite of the fact that the fluctuation
field $\phi(t)$ is exponentially decaying. What stops 
the period of efficient slow-roll inflation is the decay of the field
in the presence of a dominant quasi-constant $V(0)$.

For our set of parameters we indeed find inflation to work, and we
obtain $N_e \sim 65$, essentially the required $N_e >
60$. At this stage of our investigation we consider our results very
promising as we have not yet exploited the substantial uncertainties in
our key parameters $C_1(\mpl)$ and $\lambda(\mpl)$, which are to a
large extent determined by the \MSb input parameters $\lambda(M_H)$
and $y_t(M_t)$. The precise value of the latter, as we know, is
somewhat controversial. In any case it is remarkable that such a
scenario at worst is very close to what SM input parameters tell
us. The estimates of the uncertainties and the evaluation of the spot
in parameter space which supports our Higgs inflation scenario will be
investigated in a forthcoming analysis. Figure~\ref{fig:spectralindex}
shows how well slow-roll inflation criteria are satisfied. For our
numerical solution we obtain \mbo{\eps\approx
2.3\power{-2}} and \mbo{\eta\approx21.3\,,} when slow-roll inflation
ends at about $t\approx 450\,\tpl\,.$
\begin{figure}
\centering
\includegraphics[height=5cm]{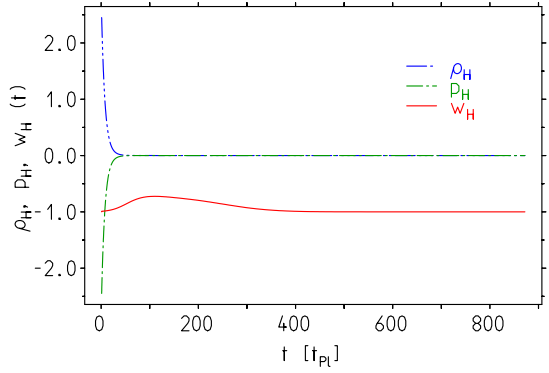}
\caption{ Higgs field density, pressure and equation of
state. The Higgs provides dark energy beyond the inflation period
which ends at about $t \simeq 450\,\tpl\,.$}
\label{fig:eqofstate} 
\end{figure}
\begin{figure}
\centering
\includegraphics[height=4cm]{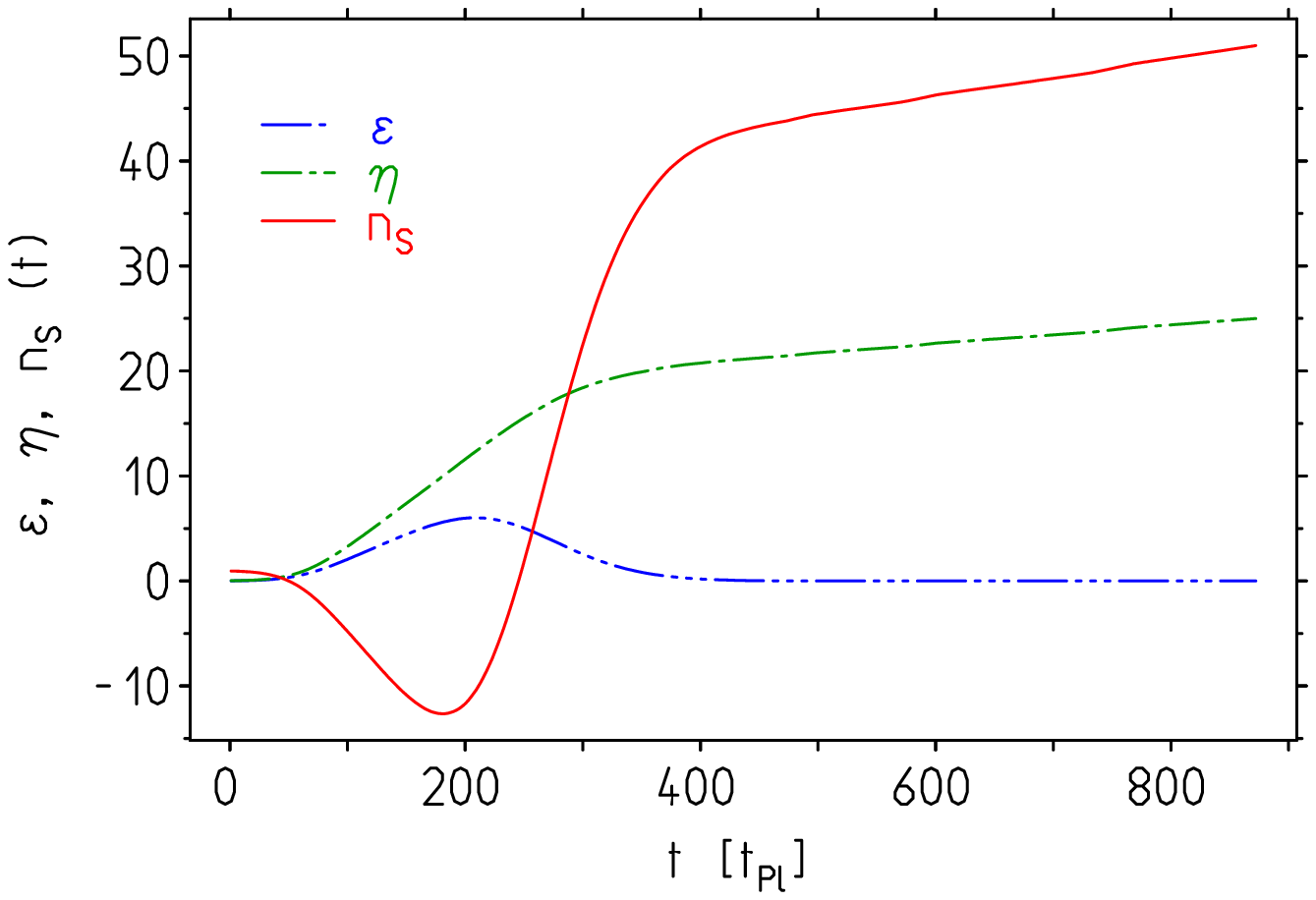}
\includegraphics[height=4cm]{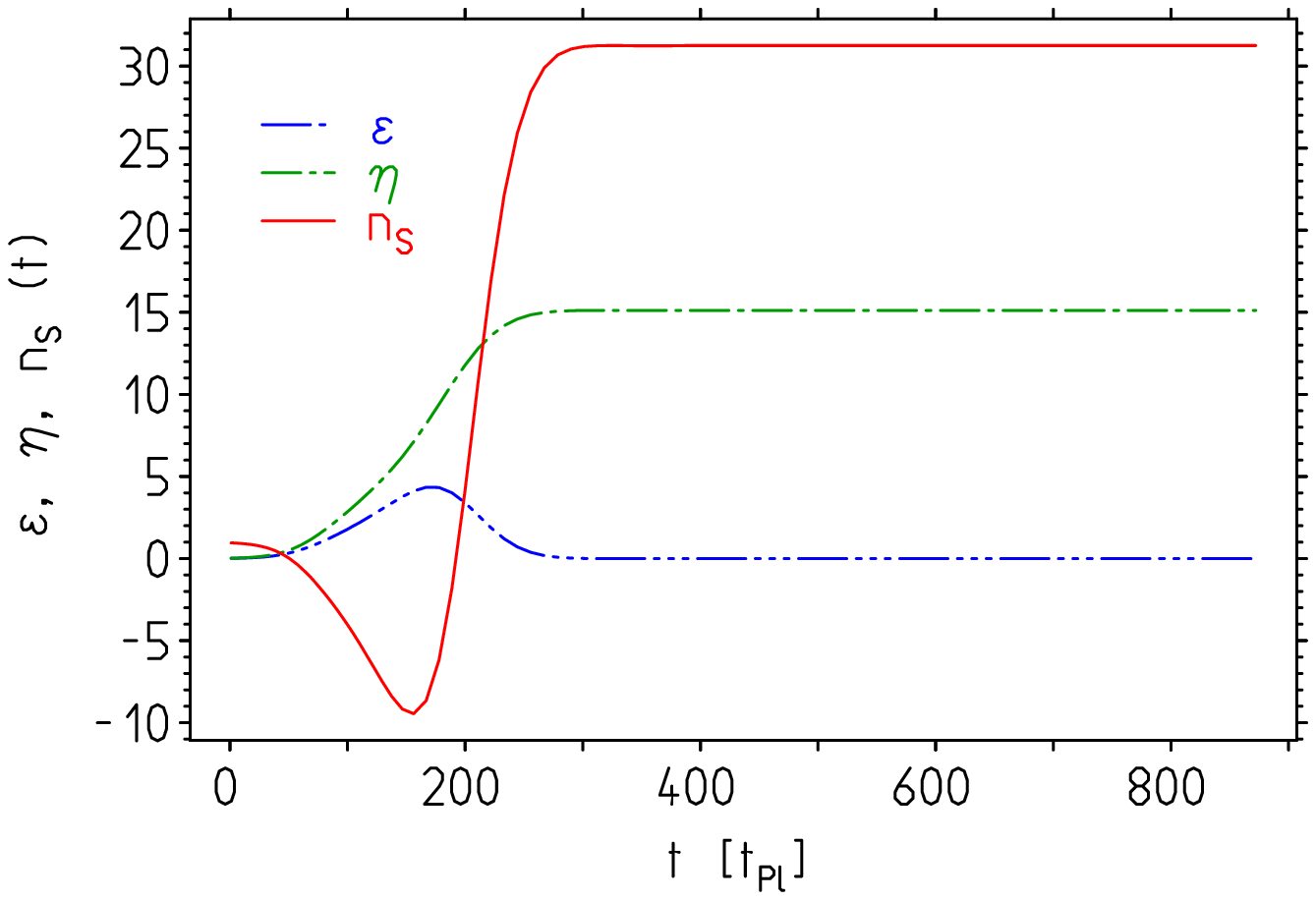}
\caption{The slow-roll coefficients $\eps$, $\eta$ and the spectral
index $n_S=1-6\eps+2\eta$ for the scalar fluctuations, as a function of time before
the bare Higgs transition point. Left: correct result taking into account the
running couplings.  Right: fake results assuming fixed couplings
as given at the Plank scale.}
\label{fig:spectralindex} 
\end{figure}
As we learn from
Fig.~\ref{fig:spectralindex} as well as by inspection of the formal SM
expressions, the indices $\eps$, $\eta$ and the resulting spectral
index $n_S=1-6\eps+2\eta$ are extremely sensitive to the effective SM
parameters, and in fact $\eta$, and consequently $n_S$, acquires values completely out of what
formal inflation requirements suggest. In fact these results depend
sensitively on the value of the Higgs field at the end of the inflation
era. How do these indices depend on the SM parameters? 

By $X$ we denote the rescaled $\phi^2$ field,
$X=\frac{\phi^2}{\Xi}=16\pi^2\frac{\phi^2}{\mpl^2}\,$, and by $\mu_m$ the unknown
renormalized mass square, $\mu_m=\frac{2m^2}{\Xi}$, in the
following. For definiteness we take $m^2=m^2_\phi$ from Eq.(\ref{mphiguess}).
This choice barely affects the results as long as we satisfy $m^2 \ll
\delta m^2\,.$ For the Higgs potential we obtain
\bea
\eps=16\pi X \left[\frac{3\,(\mu_m+C+\lambda)+\lambda
X}{3\,(2\,(\mu_m+C)+\lambda)+6\,(\mu_m+C+\lambda)\,X+\lambda X^2}\right]^2
\eea
such that for $\phi^2 \to 0$ (late inflation) 
\bea
\eps=16 \pi X\,\left[\frac{\mu_m+C+\lambda}{(2\,(\mu_m+C)+\lambda)}\right]^2\approx \left\{ \begin{tabular}{ccc}
$64\pi^3\,\frac{\phi^2}{\mpl^2}$ & for & $\mu_m+C\gg \lambda$\\
$256\pi^3\,\frac{\phi^2}{\mpl^2}$ & for & $\mu_m+C\ll\lambda$
\end{tabular}\right. \epo
\eea
At inflation begin $X\gg \mu_m+C,\lambda$ ($\phi^2$ large) we have
\ba
\eps \sim \frac{1}{\pi}\,\frac{\mpl^2}{\phi^2}\epo
\ea
It is obvious that during inflation $\eps$ is naturally
expected to be small and in the limiting cases is independent of any
SM parameters.

The second index 
\bea
\eta=24\pi\,\frac{\mu_m+C+\lambda+\lambda
X}{3\,(2\,(\mu_m+C)+\lambda)+6\,(\mu_m+C+\lambda)\,X+\lambda X^2}
\eea
behaves differently. For $\phi^2 \to 0$ (late inflation) 
\bea
\eta =8\pi\, \frac{\mu_m+C+\lambda}{2\,(\mu_m+C)+\lambda}\approx \left\{ \begin{tabular}{ccc}
$4\pi$ & for & $\mu_m+C\gg \lambda$\\
$8\pi$ & for & $\mu_m+C\ll\lambda$
\end{tabular}\right. \epo
\eea 
So the SM predicts very large values when the field gets small towards
the end of inflation. If an inflation criterion $\eta \ll 1$, or even
$\eta <3\eps$ in case we require $n_S=1-6\eps+2\eta<1$, would be a true
necessary condition this would rule out the SM Higgs as a inflaton.
At the beginning of inflation $X\gg C,\lambda $ ($\phi^2$ large) we have
\ba
\eta \sim  \frac{3}{2\pi}\,\frac{\mpl^2}{\phi^2}\epo
\ea 
A small $\phi^2$ makes $\eps\sim \phi^2/\mpl^2$ in any case small,
while a small $\eta$ requires $\phi$ not too small, as for $\phi\to 0$
$\eta$ takes large values depending on whether the mass or the
interaction term of the potential dominates. As soon as one of the
terms in the Lagrangian dominates, inflation is insensitive to $m^2$ or
$\lambda$, respectively, as a rescaling of the potential is not
affecting inflation. When $\lambda \to 0\,:$ $\eta\to 4\pi\,$; when
$m^2\to 0\,:$ $\eta \to 8\pi\,$. This is what we observe in
Fig.~\ref{fig:spectralindex}. Note that these results seem to be
universal for a $\phi^4$ scalar potential and insofar are not specific
for the Higgs sector with its particular parameters. However, at look
at Eqs.~(\ref{phidecay}) and (\ref{phidecay2}) tells us that SM parameters have a strong impact
on the decay rate of the scalar fluctuation field. The fact that with
our educated guess for the initial value $\phi_0$ at $\mpl$ we obtain
$\eps$ very small while $\eta$ is obtained too large is a direct
consequence of the fact that the field $\phi_e$ at late inflation
times is obtained to be so small, which, however, strongly depends on
the values of $m$ and/or $\lambda$ and hence on specific inputs
$\lambda(M_H)$ and $y_t(M_t)$. Because of the strong decay of the
field an increase of $\phi_0$ essentially does not affect $\eta$ at
late inflaton times. This does not imply that we cannot get a
sufficiently large inflation factor $N_e$, fortunately. The problem
appears to be the spectral index for the scalar perturbations
$n_S=1-6\eps+2\eta$, to be considered next, which is constrained by observation to
be $n_S<1$ requiring $\eta \leq 3\eps$, which seems to be very hard to
satisfy for any symmetric Higgs type potential.

\begin{figure}
\centering
\includegraphics[height=4cm]{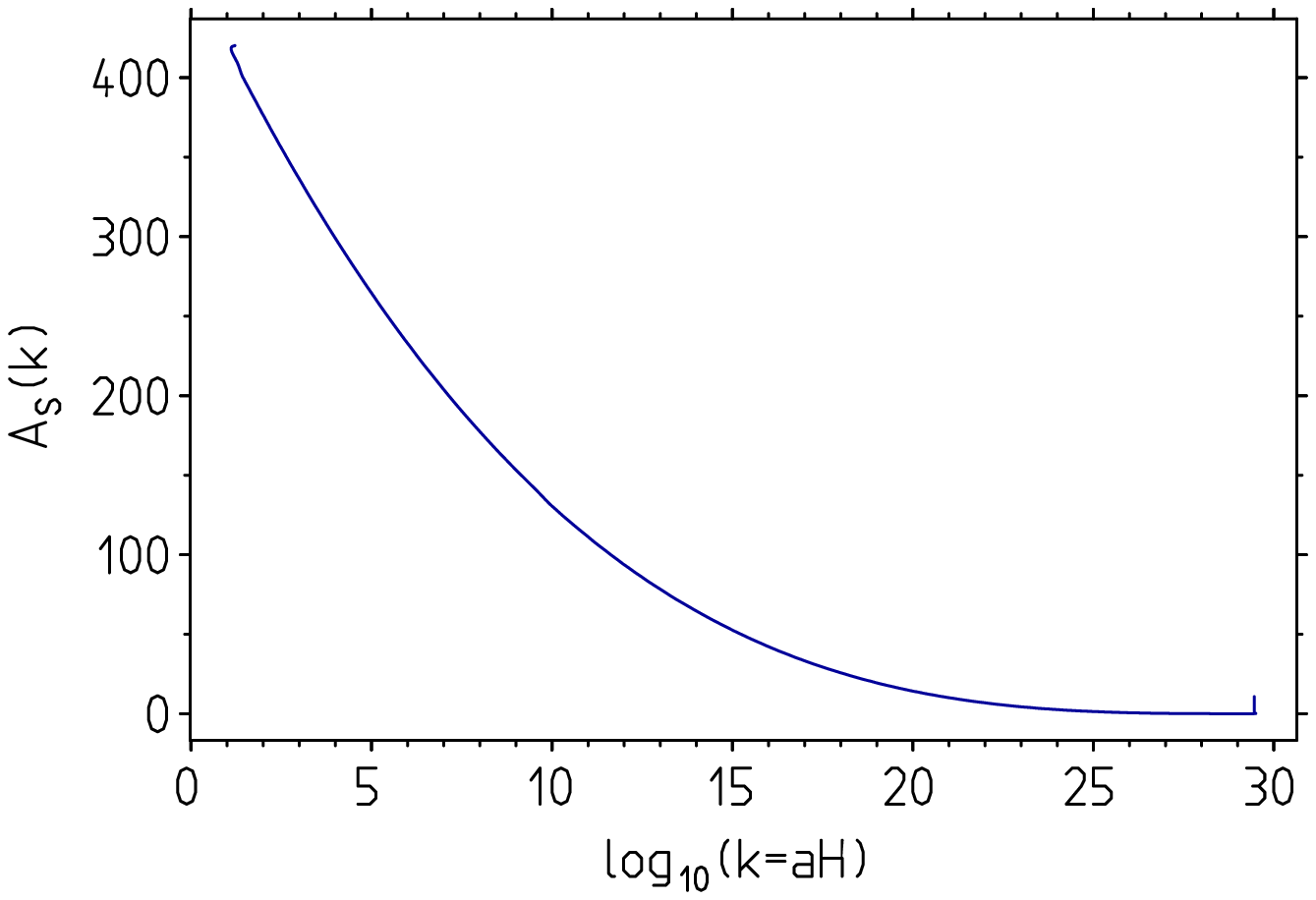}
\includegraphics[height=4cm]{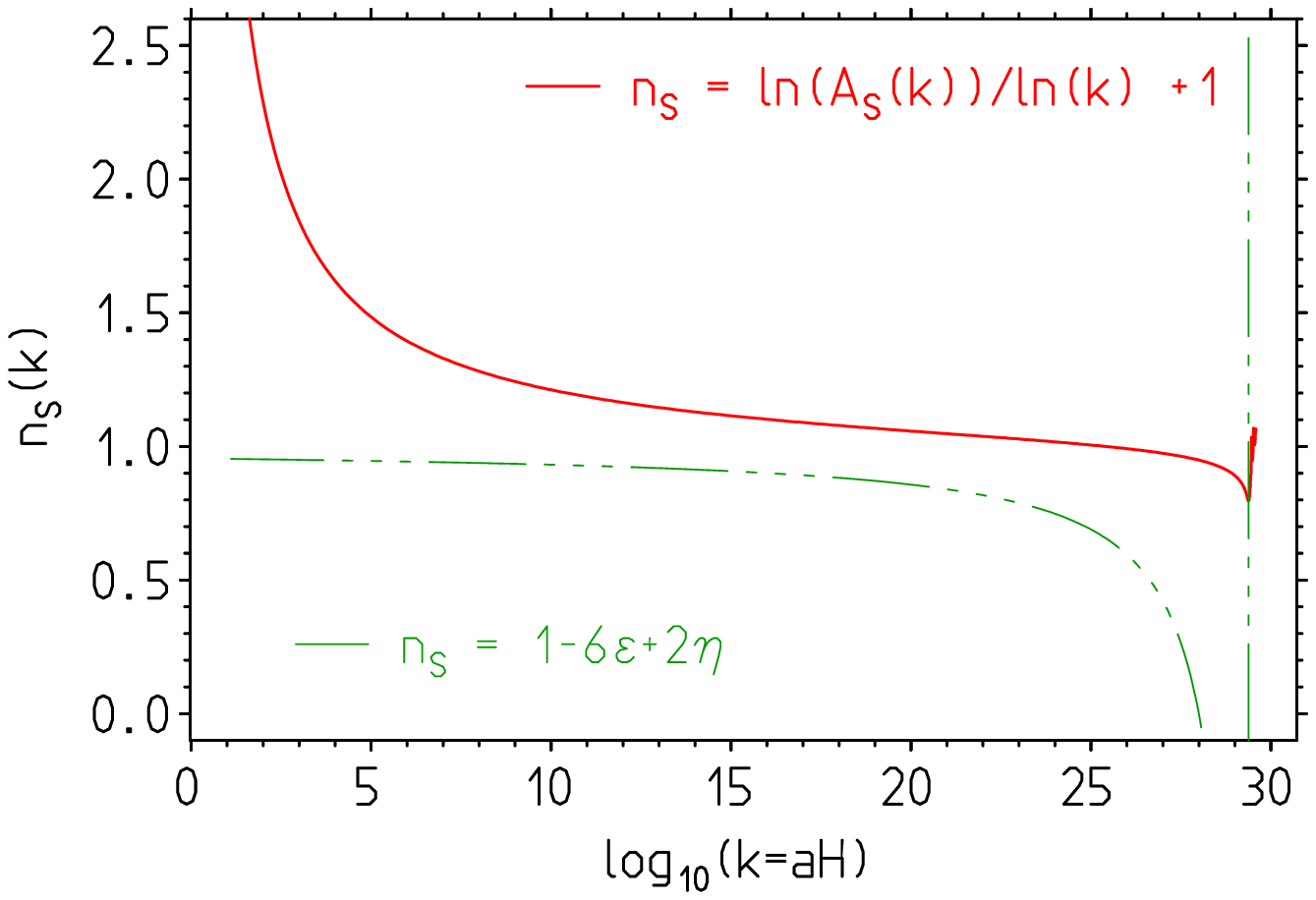}
\caption{The spectral amplitude $A_S(k)$ as a function of $k=aH$
[left] and a comparison of the indices $n_S=\ln A_S(k)/\ln k+1$ versus
$n_S=1-6\eps+2\eta$ [right].}
\label{fig:Ask} 
\end{figure}

\begin{figure}
\centering
\includegraphics[height=4cm]{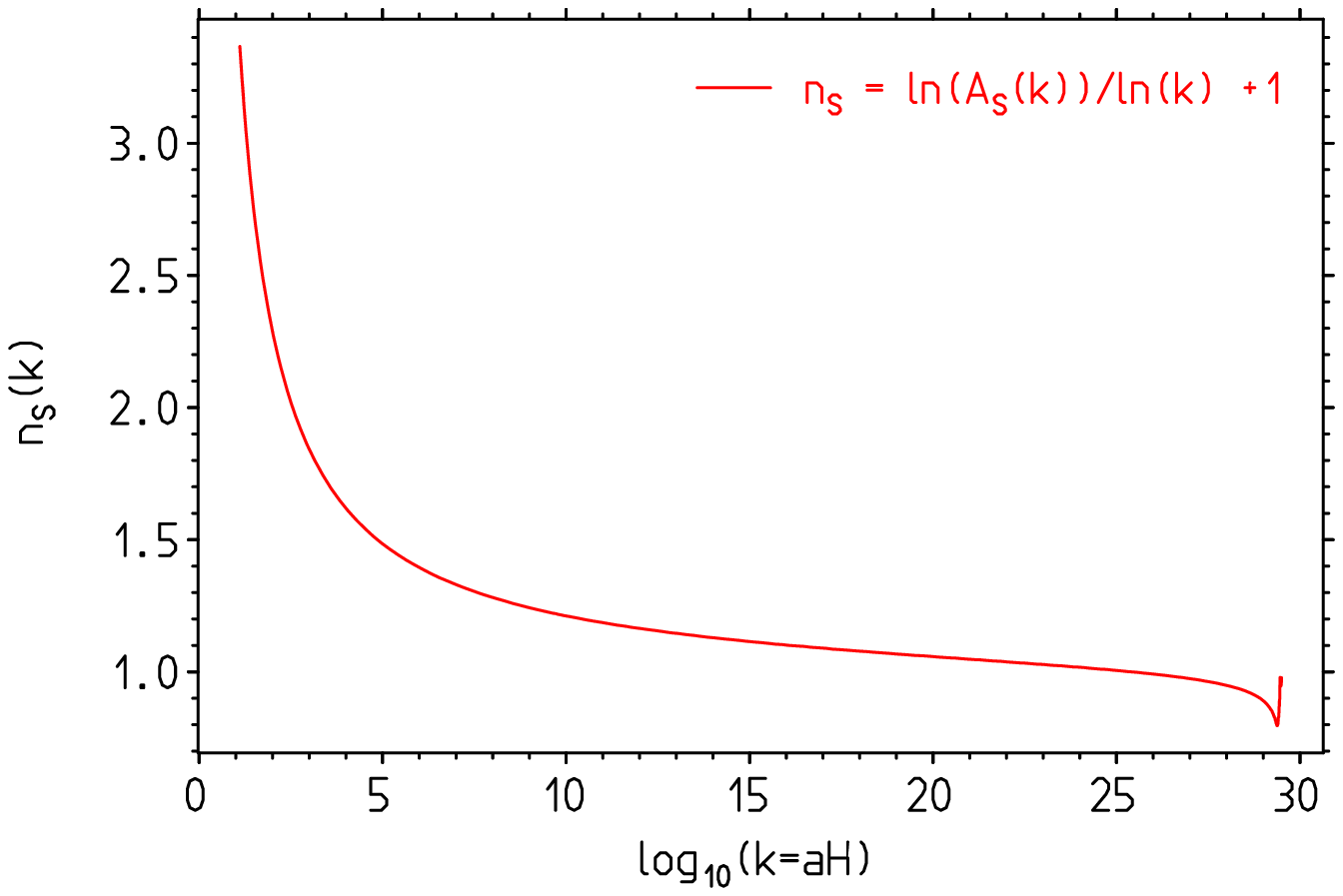}
\includegraphics[height=4cm]{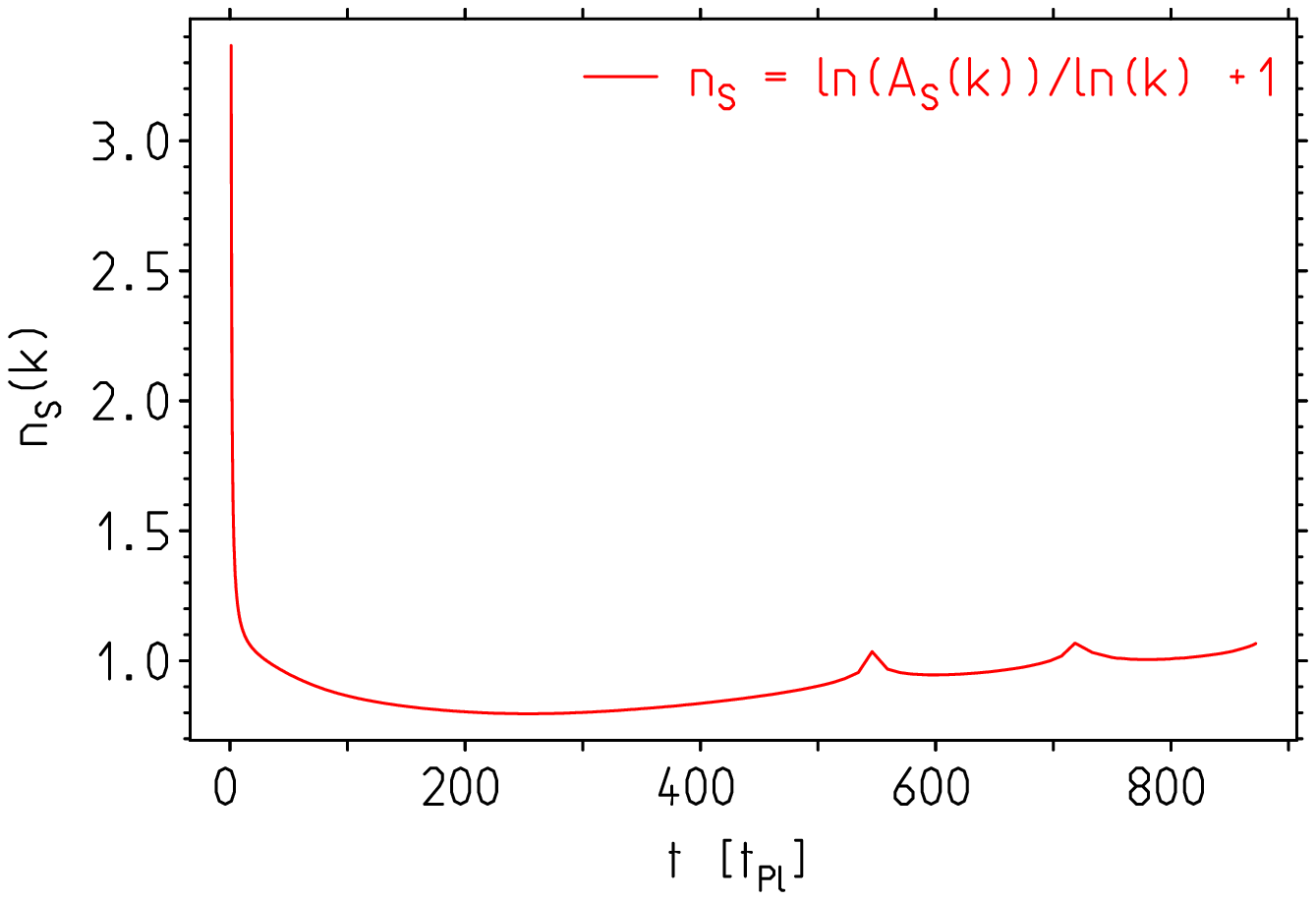}
\caption{The index $n_S(k)$ extracted from $A_S(k)$ as a function of
$k=aH$ [left] and as a function of time $t$ in units of $\tpl$.}
\label{fig:nSk} 
\end{figure}

The indices just considered play a role in estimates of the scalar
density fluctuations \mbo{\delta \rho=\frac{\D V}{\D
\phi}\,\delta\phi\,,} which are tailored by
inflation~\cite{Mukhanov:1981xt,Mukhanov:1985rz,Mukhanov:1990me} and
exhibit a spectrum
\bea
A_S(k)
&=&\left.\frac{128\pi}{3}\,\frac{V^3}{\mpl^6\,(V')^2}\right|_{k=aH} \\
&=&\frac{\pi}{9\,(16\pi^2)^3}
\frac{\left(3\,(2\,(\mu_m+C)+\lambda)+6\,(\mu_m+C+\lambda)\,X+\lambda\,X^2\right)^3}{
X\,\left(3\,(\mu_m+C+\lambda)+\lambda X\right)^2}\nn
\label{Ask}
\eea
to be evaluated at the moment when the physical scale of the
perturbation \mbo{\lambda=a/k} is equal to the Hubble radius
\mbo{H^{-1}\,} and thus at the event horizon.
For small fields $X \to 0$ we have
\ba
A_S(k) &\sim& \frac{\pi}{3\,(16\pi^2)^3}
\frac{1}{X} \frac{(2\,(\mu_m+C)+\lambda)^3}{(\mu_m+C+\lambda)^2}\nn\\ &\sim& \left\{ \begin{tabular}{c}
$\frac{8\pi}{3\,(16\pi^2)^4}\,(\mu_m+C)\,\frac{\mpl^2}{\phi^2}\,;\,\, \mu_m+C \gg \lambda$\\
$\frac{\pi}{3\,(16\pi^2)^4}\,\lambda\,\frac{\mpl^2}{\phi^2}\semis \hspace*{1.2cm} \mu_m+C \ll \lambda$
\end{tabular}\right.
\ea
while for large fields $X \gg C,\lambda$ we find the behavior
\ba
A_S(k)\sim 
\frac{\pi}{9}\,\lambda\,\frac{\phi^6}{\mpl^6}\epo
\ea
Observations are parametrized by a power spectrum
\mbo{A_S(k)\propto k^{n_S-1}} where
\mbo{n_S=1-6\eps+2\eta\,.}
The last relation is obtained by assuming that the Higgs potential
dominates the scene and that a change of $k$ is given solely by a change of $\phi$:
\bea
\frac{\D}{\D \ln k}=-\frac{V'}{2H^2}\frac{\D}{\D \phi}
\eea 
in the slow-roll limit. The latter relation explains how $V''$ and
herewith $\eta$ come into play. However, for the SM, given the fairly
large quasi cosmological constant $V(0)$, which is largely
determining the Hubble constant $H$ during inflation, the above
relation and therefore the relation $n_S=1-6\eps+2\eta\,$ does not
apply. We therefore prefer to extract $n_S$ from the amplitude
$A_S(k)$ (\ref{Ask}) directly (see Fig.~\ref{fig:Ask}) and calculate
\bea
n_S(k)=\frac{\ln A_S(k)}{\ln k}+1\epo
\label{nSk}
\eea
We find the result displayed in Fig.~\ref{fig:nSk} which differs substantially from the $n_S=1-6\eps+2\eta$
as illustrated in Fig.~\ref{fig:Ask} [right].
Actually $n_S(k)$ obtained via the exact definition provides results
much closer to the reported observational value. 

In our LEESM scenario the CMB temperature fluctuations would be
directly related to the Higgs fluctuation field and CMB data would be
a first direct access to measure the Higgs wave function in the early
universe. With the above estimates we obtain
\mbo{n_S\approx 1.067} for $t\approx\,870\,\tpl$ and $n_S\approx 0.866$
at $t\approx\,450\,\tpl\,,$ which corresponds to the end of slow-roll
inflation. These results
confront with the recent Planck mission result~\cite{PlanckResults}
\mbo{n_S=0.9603\pm0.0073\,.} I have not yet estimated the uncertainty,
which however is expected to be large enough not to be in plain
conflict with the data. More importantly, as we learn form
Fig.~\ref{fig:nSk}, $n_S$ extracted via (\ref{nSk}) is moderately
below unity before it reaches values slightly above unity when we
approach $\mu_0$. In comparing ``theory with experiment'' we should be
aware that the observed spectrum is extracted over a cut $k$-range in
the tail $k_{\rm min}<k$ of the amplitude $A_S(k)$ where the signal is
not buried in the noise, which means $k< k_{\rm max}$. This leads to a
much better agreement, but requires to take into account the cuts
adopted in extracting the index from the CMB data.

The key lesson here is that, assuming $C(\mu)$ and $\lambda(\mu)$
being given, the spectral index is a monitor for the value of the
otherwise unknown Higgs field $\phi(\mu)$, where $\mu$ is the scale
accessible to observation. Once we kow $\phi(\mu)$ at some $\mu$ we
can determine $\phi(\mpl)$ at the Planck scale, by solving the
evolution equations from $\mu$ to $\mpl\,.$

Our analysis is a first estimate of the gross features of this Higgs
inflation scenario. Here the SM Higgs field (actually four real heavy
scalar fields of equal mass) is the inflaton scalar field, in the
symmetric phase of the SM. In SM inflation the couplings of the
inflaton to other particles are known, which is important for both the
dynamics of inflation and reheating. As inflation pushes the system
out of equilibrium through dramatic cooling, the effective couplings
could change as much such that inflation gets stopped premature. The
following reheating could reverse the process, such that the sign of
the Higgs potential mass term could actually alternate for some short
time interval. A detailed study of relaxation times of processes
participating is required here. The question is whether at the Planck
scale we have a system in thermal equilibrium. Then it could well be
that inflation and the subsequent Higgs mechanism go so fast that the
screening and antiscreening processes are slow enough such that an
oscillatory inflation-reheating era is avoided. In our evaluation
above we have assumed that the couplings follow as in thermal
equilibrium and reheating has not yet been taken into account.

Concerning the running of couplings, relevant are the virtual
processes which are dressing the top quark Yukawa coupling in
$Ht\bar{t}$ and the Higgs self-coupling in $HH\to HH,WW,ZZ$ as well as
those relevant for the gauge couplings. At first sight a substantial
change of the couplings between $\mu_0$ and $\mpl$ is not expected if
we look at Table~\ref{tab:params}, however, this is not necessarily
true as the $\beta$-functions utilized so far do not take into account
that above $\mu_0$ the Higgses are very heavy. Above $\mu_0$ the four
heavy Higgses $H,\phi,\phi^\pm$ do not contribute any more to the
coefficients of the $\beta$-functions and the corresponding scalar
loop contributions should be dropped. The corresponding changes have
to be worked out yet. Besides the gauge boson self-couplings only the
$H,\phi \to f\bar{f}$, the $\phi^\pm \to f\bar{f}'$ and the quartic
$HH\to WW,ZZ$, $\phi^+\phi^-\to \gamma \gamma$, $\cdots$ type
couplings are effective. The latter all include two heavy Higgs fields
and two gauge boson fields and are barely effective
in this phase. All $H\to WW$ single Higgs to two gauge bosons are
absent in the symmetric phase (in the broken phase they are induced by
the Higgs mechanism and proportional to the Higgs VEV $v$), such that
they are not effective in Higgs decays. A detailed analysis is beyond
the scope of the present work. One point is clear however, the
seemingly minor differences in the running parameters between $\mu'_0$
and $\mpl$ have a dramatic impact on the value of the coefficient
$C'(\mu)$ of the quadratic divergence, namely, the latter is zero at
$\mu'_0$, while it is magnified by the huge factor $\mpl^2$ away from
$\mu'_0$. It is responsible for the Higgs phase transition on the
one hand as well as for the large Higgs masses in the symmetric phase
above $\mu'_0$ on the other hand. The running of the couplings in the
symmetric phase thus has a dramatic impact on the specific properties
of inflation.

\section{Comment on reheating and baryogenesis}
The Higgses near the Planck scale have an effective mass about $m_{H0}
\simeq 3.6\power{17}~\gv$ and thus can be produces in processes like
$WW\to HH$ or $t\bar{t} \to H$ at times before the phase transition
takes place provided the temperature is above the corresponding Higgs
mass thresholds $2 m_{H0}$ $\sim 8.4\power{30}~\degK$ or $ m_{H0}$
$\sim 4.2 \power{30}~\degK$, respectively. This may not be too
relevant as the heavy Higgses are \textbf{primordial fields} of the Planck
medium. A Higgs there has a width dominated by $H\to t\bar{t}$ decay
with
\bea
\Gamma_H&\simeq& \Gamma(H\to t \bar{t})\simeq\frac{m_{H0}}{16 \pi}\,N_c\,y^2_t(\mpl)\nn\\
&\simeq& 7.35\power{-3}\,m_{H0}\simeq 2.65\power{15}~\gv
\eea
which yields a life time
\bea
\tau_H=1/\Gamma_H\simeq 2.5\power{-40}~\mbox{sec}\,,
\eea
which actually, for this process, supports the argument that, for some
time, the effective couplings essentially do not change when the system
is driven out of equilibrium.  Here we used our result
$y_t(\mpl)=0.3510$. We note that $\tau_H$ is large in terms of the
Planck time $\tpl \simeq 5.4
\power{-44}~\mbox{sec}$ as well as the bare Higgs transition time
$t_H\approx 4.7\power{-41}~\mbox{sec}$ and the effective Higgs
transition time $t'_H\approx 8.5\power{-40}~\mbox{sec}\,.$ The drop of
the $V(0)$  happens at $t_{\rm CC}\approx 2.1\power{-40}~\mbox{sec}\,,$
halfway between $t_H$ and $t'_H\,.$  

The SM predicts that the Higgses produce top--anti-top quarks most
abundantly. In addition we can say that $\Gamma_H
\ll H(t)=\dot{a}(t)/a(t)$ as $H(t) \geq 7.2\power{16}~\gv$ during inflation,
and actually until the drop of the CC discussed below.
The heavy Higgses represent decaying non-relativistic matter such that
their density scales with time as
\bea
\rho_\phi(t)=\rho_\phi(t_i) \left(a_i/a(t)\right)^3 \,\E^{-\Gamma_H\,(t-t_i)}\,,
\label{HeavyHiggsMatter}
\eea
as a solution of $\dot{\rho}_\phi+3H\rho_\phi+\Gamma_H \rho_\phi=0\epo$
The energy density of top/anti-top quarks produced by the Higgs decays
satisfies the conservation equation (see e.g. Ref.~\cite{Weinberg:2008zzc})
\bea
\dot{\rho}_t +3\,H\,(\rho_t+p_t) = \Gamma_H\,\rho_\phi
\eea
with $H^2=\ell^2\,\left(\rho_\phi+\rho_t+\cdots\right)\,,$
and since the top quarks are relativistic (in the symmetric phase) $p_t=\rho_t/3$
such that
\bea
\rho_t&=&\Gamma_H\,\rho_\phi(t_i)\,a^3(t_i)/a^4(t)\,\int_{t_i}^t\,\D t'\, a(t')\,\E^{-\Gamma_H\,(t'-t_i)}\\
 &\leq&\Gamma_H\,\rho_\phi(t_i)\,a^3(t_i)/a^4(t)\,\int_{t_i}^t\,\D t'\, a(t')\epo
\eea
At these times the energy density is still dominated by the inflaton,
such that $a(t)=a(t_i)\,(t/t_i)^{2/3}$ and hence~\cite{Weinberg:2008zzc}
\bea
\rho_t&=&\frac{3}{5}\,t_i\,\Gamma_H\,\rho_\phi(t_i)\,\left(\frac{t_i}{t}\right)^{8/3}\,\left(\left(\frac{t}{t_i}\right)^{5/3}-1\right)\epo
\eea
The maximum is reached for $t=(8/3)^{3/5}\,t_i$ with
$H(t_i)=2/3\,t_i$, still well consistent with the assumption $\Gamma_H \ll
H(t_i)$. The maximum matter density is then constrained by the bound
\bea
\rho_{t\,{\rm max}}&\leq&(3/8)^{8/5}\,t_i\,\Gamma_H\,\rho_\phi(t_i)
=0.139\,\left(\Gamma_H/H(t_i)\right)\,\rho_\phi(t_i) \nn \\
&\simeq&
0.139\,\frac{3\sqrt{3}\,y^2_t(\mpl)}{64\,\pi \sqrt{\pi}}\,\mpl^3\,m_{H0}
\simeq
1.6\power{71}~\gv^4\epo
\label{topmatter}
\eea
In fact, the numerical estimate for the true maximum yields
$\rho_{t\,{\rm max}}\simeq 1.2\power{71}~\gv^4$ reached at $t\simeq
1.74\,\tpl\,,$ well within the estimated bound. The main difference is
due to taking properly into account the Higgs width. The evolution of
$\rho_t$ is also displayed in Fig.~\ref{fig:rhooft}. These values
compare to the cosmological constant vacuum density
\bea
\rho_\phi 
&=& V(0)+\frac{m^2(\mu_0)}{2}\phi_e^2+\frac{\lambda(\mu_0)}{24}\phi_e^4
\approx V(0) \nn\\&\simeq& 3.5\power{-6}\,\mpl^4\simeq
7.67\power{70}~\gv^4\epo
\eea
After the produced $t\bar{t}$ pairs have been interacting sufficiently
often
to thermalize, the top flavored medium has a temperature $\rho_t=g_*
\pi^2 T^4/30$ and the maximum reheating temperature reaches
\bea
T_{\rm max}=\left(\rho_{t\,{\rm
max}}/\left(g_*\,\pi^2/30\right)\right)^{1/4}\simeq 8.9\power{30}~\degK\,,
\eea 
with $g_*=12 \frac78$ for top quarks. As reheating temperature one
defines~\cite{Kolb:1990vq}
\bea
T_\mathrm{RH}&\equiv& T(t=\tau_H)\simeq
0.55\,{g_*}^{-1/4}\,\left(\Gamma_H \mpl \right)^{1/2}\nn\\&\approx&
1.5\power{-2} \mpl\simeq 2.1\power{30}~\degK\,,
\label{TRH}
\eea
usually considered to be the begin of the radiation dominated phase of the universe.

The physical ``charged'' channels $H^+\to t\bar{b}$ and $H^-\to
b\bar{t}$ have by a factor $y_b/y_t$ reduced rates and $H\to b\bar{b}$
is reduced by $(y_b/y_t)^2$. Very important are the ``charged'' decays $H^+\to
t\bar{d},u\bar{b}$ and $H^- \to d\bar{t},b\bar{u}$ which exhibit the dominant
complex CP-violating CKM~\cite{CKM} matrix-elements\footnote{Present data~\cite{pdg} yield $V_{td}\sim A\lambda^3\,\left(1-\rho-\I \eta\right)$
and $V_{ub}\sim A\lambda^3\,\left(\rho-\I \eta\right)$ with
$\lambda=0.22535\pm0.00065$, $A=0.811^{+0.022}_{-0.012}$,
$\rho=0.131^{+0.026}_{-0.013}$ and the CP violating phase $\eta=0.345^{+0.013}_{-0.014}$.
The overall CP violation is characterized by the Jarlskog invariant
$J=\lambda^6\,A^2\,\eta\approx 2.97 \power{-5}$ and the Jarlskog determinant $\delta_J=J\,(y_t^2-y_c^2)\,
(y_t^2-y_u^2)\,(y_c^2-y_u^2)\,(y_b^2-y_s^2)\,(y_b^2-y_d^2)\,(y_s^2-y_d^2)\,\approx
9.94\power{-24}$ a tiny number, which however together with a large
Higgs induced dark energy term, decaying predominantly into heavy
flavors, may lead to realistic magnitude for the
baryon-asymmetry.}. Also the subsequent processes $t
\to W d$ or $b \to W u$ exhibit the same CKM couplings, which are able to
contribute to the baryon-asymmetry. 

In standard baryogenesis scenarios some unknown heavy particle $X$,
usually assumed to have been pair-created in the hot early universe,
when $k_BT$ exceeded all particle masses, are assumed to decay into
pairs of particles of different baryon and lepton number. This can
produce a baryon- or lepton-asymmetry, respectively, if $C$ and $CP$
are violated as they are in the SM. The necessary $B$ violation are
assumed to be produced by appropriate dimension 6 four-fermion
operators~\cite{Weinberg:1979sa,Buchmuller:1985jz,Grzadkowski:2010es}.
In our LEESM scenario, the latter are naturally expected as
$(E/\Lambda_{\rm Pl})^2$ terms in the low energy expansion. At the
scale of the EW phase transition the Planck suppression factor is
$1.3\power{-6}$. The basic asymmetry parameter is
$\epsilon=B_X+B_{\bar{X}}=r-\bar{r}$, where $B$ are the baryon numbers
produced and $r$ and $\bar{r}$ are branching fractions, which would be
equal ($\bar{r}=r$) if $C$ and/or $CP$ would be
conserved~\cite{Kolb:1990vq}. One assumes that the particle numbers in
thermal equilibrium agree $n_X=n_{\bar{X}} \sim n_\gamma$ and the
baryon number density in units of the entropy density is given by
$n_B/s\sim \epsilon\, n_X/g_* n_\gamma \sim \epsilon/g_*$. Here $g_*$
is the number of relativistic degrees of freedom produced in the decays. Typical
values are $\epsilon\sim 10^{-8}$, sufficient to reproduce the
observed baryon asymmetry $\eta_{_{B}}\sim 10^{-10}$.

When $k_BT(t)>M_X$ and $H(t)< \Gamma_X$, $X$-creation
and $X$-annihilation are equally efficient and the expansion rate is
slow enough such that there is sufficient time for the system to stay
in thermal equilibrium. No net asymmetry can develop in this
case. During inflation, until the drop of the CC at $t_{\rm CC}\approx
2.1\power{-40}~\mathrm{sec}\,,$ we actually have a large Hubble
constant $H=\dot{a}/a (t) \gg \Gamma_X$. In this case, as long as
$k_BT(t)>M_X$ still $X$ production is effective and the various
radiative components still follow the equilibrium distributions with
rescaled temperature $T(t)\propto 1/a(t)$ and the system behaves
as if in thermal equilibrium. So, what we need is $H=\dot{a}/a (t)
\gapprox \Gamma_X$ and $k_BT(t)<M_X$ in which case the inverse decay
is blocked and the system is truly out of equilibrium and the
out-of-equilibrium condition for baryogenesis is satisfied. Our
scenario is somewhat different. Our $X$ particle are the four very
heavy primordial Higgses from the Planck medium, the properties of
which we know, in particular their masses, widths and branching
fractions, as well as their $C$ and $CP$ violating coupling
structure. The system seems to be out of equilibrium way down to the
Higgs transition point $\mu'_0$, and the immediately following EW
phase transition, which is of pronounced 1st order type triggered by
the sing change in the effective mass ${m'}^2$. In our case, Higgs
reproduction gets stopped somewhat earlier by the drop of the CC (see
Fig.~\ref{fig:Lagrangianext}). The Higgs width there is about
$\Gamma_H\approx 1.39 \power{-4}\,\mpl$. Before, the drop of the CC at
time $t\approx 3796\,\tpl$ the Hubble constant is $H\approx 2.03
\power{-4}\,\mpl\,,$ while after the drop, at $t\approx 3810\,\tpl$,
$H$ has decreased suddenly to $H\approx 5.50\power{-12}\,\mpl$ and
Higgs recreation is stopped. At this stage the inflaton field is
converted completely into radiation. A detailed analysis and numerical
estimate of the baryon asymmetry which one can obtain in this way 
is missing at this point. If we adopt the analysis elaborated in Ref.~\cite{Kolb:1990vq}
as the way-out-of-equilibrium scenario, which covers our Higgs
inflation scenario, the baryon asymmetry is given by $n_B/s\simeq
\epsilon\, T_\mathrm{RH}/m_\phi$, with reheating temperature
$T_\mathrm{RH}$ given in Eq.~(\ref{TRH}). Then, assuming $\epsilon
\sim 10^{-8}$ one estimates $n_B/s \simeq 5.0\power{-9}$. What remains
to be worked out is the basic $B$ violation parameter $\epsilon$. The
value adopted looks very plausible in our scenario but depends on
unknown couplings accompanying the $B$ number violating dimension 6
operators.

It is interesting to see that the large cosmological constant energy
density is converted predominantly into, yet massless, top flavored
matter/antimatter, which is known to undergo matter-antimatter
annihilation.  It seems that we are all descendants from a dense
top--anti-top sea, which cascades down to the light quark world we
live in. Thereby we have to undergo all CKM physics.  During the EW
phase transition the particles acquire their mass and heavier
particles start to decay into lighter ones. The effective mass
hierarchy mix-up, illustrated by Fig.~2 of
Ref.~\cite{Jegerlehner:2013cta}, likely plays a role here.  Close to
the phase transition the effective top-quark mass is not yet clearly
heavier than the effective $W$-boson mass, for example. So top quarks
could be quasi-stable and form toponium states for some time after the
EW phase transition has taken place, and before they decay via $t\to
Wb$ and then cascade down to the light quarks. It is interesting to
note that in the symmetric phase those fermion modes win which have
the strongest Yukawa coupling, i.e. the top quark flavor. After the EW
phase transition, because of the mass-coupling relation, when the
universe cools further down, the heavier particles decay into lighter
ones. Now the modes with the weakest Yukawa couplings win, as they
have larger phase space, and survive in form of normal matter. This
also may shed some light on the question about the huge mass hierarchy
between the heaviest and the lightest quark $M_t/m_u\sim 10^{5}$,
which in our scenario must be large enough to get a sufficient amount
of normal light matter. Whether this intricate EW phase transition
scenario leaves room for dark matter relicts is an open problem.

It is important also to remind us that it is the EW phase
transition\footnote{Because of the finite temperature effect the EW
phase transition takes place always after the Higgs mechanism.}  which
puts the SM into operation with all its properties we are familiar
with. QED, in particular, with its special role in the development of
structure during the following evolution of the universe comes into
existence only with the EW phase transition. Normal photon radiation,
the photon as the only massless particle, particles of definite
charge, matter-antimatter annihilation into light, Dirac fields and all that show up
in the broken phase only. Particularly interesting is what happens
with the most abundantly produced top quarks, which decay into the
lighter flavors. It is interesting that this proceeds through the
$b$--quark sector, which exhibits the large component of CKM
CP--violating phase. Since, in our scenario, the EW phase transition
is carried along by the Higgs transition, the system likely would
intermittently be far from equilibrium. Both, are key ingredient for
the explanation of the baryon-asymmetry and for understanding
baryogenesis. So, possibly, in this scenario the origin of the
baryon-asymmetry could well have its explanation within the SM.

One of the main consequences of our LEESM scenario is that the SM
hierarchy problem is not a problem of the SM but the solution for
inflation and dark energy in the early universe (see
Ref.~\cite{Jegerlehner:2013nna} for more details). Since the Higgs
VEV $v(\mu^2)$ emerges as a low energy phenomenon (order parameter)
from the phase transition regime at or near $\mu'_0$, where the
quadratic divergence is nullified, there is no hierarchy problem in
the broken phase since as a result of the well known mass coupling
relations all masses, including the Higgs mass itself, have values
proportional to the EW scale $v$, up to factors essentially given by
the SM couplings, which in any case depend logarithmically on the
scale only.

\section{The cosmological constant}
\label{Sect:CC}
The cosmological constant problem (see
e.g. Ref.~\cite{Straumann:1999ia}) has been a persisting problem for
a long time already. It usually is considered to be the most severe
fine tuning problem within the SM\footnote{As emphasized in
Ref.~\cite{Volovik:2005zu,Volovik:2006bh}, one should note that in
condensed matter systems the macroscopic ground state energy density
is not determined by the quantum fluctuations but rather by
temperature and pressure of the system which are determined by the
environment which can change with time.}. As we have seen, the SM
predicts a huge time-dependent CC, at $\mpl$ equivalent to
$\rho_\phi\simeq V(\phi)
\sim 2.77\,\mpl^4\sim 6.13\power{76}~\gv^4$, for the given initial
field value (\ref{phinullplus}), while the value observed today is
$\rho_{\rm vac}=\mu_\Lambda^4$ with $\mu_\Lambda\sim
0.002~\mbox{eV}$! In the unbroken phase the CC is essentially
provided by the quartically enhanced Higgs potential $V(\phi)$, which
identifies the CC as a field, however, with a weakly scale dependent
vacuum contribution $V(0)$, which shortly after the begin of inflation
starts to be dominating. As we already know, later, the Higgs
mechanism contributes $-\frac{\lambda}{24}v^4\sim 1\power{9}~\gv^4$ to $V(0)$ 
via the non-vanishing VEV $v$. Although much smaller than $V(0)$ at scales
$\mu>\mu_{\rm CC}$, it represents a large negative
contribution to the vacuum energy density\footnote{At scale $\mu_0 \sim 1.4
\power{16}$ we have $\lambda \sim 0.1393$ and $v
\sim 638~\gv$ such that $\frac{\lambda}{24}\, v^4\sim
9.6\power{8}~\gv^4$. Converted with the factor $\kappa=8\,\pi\,G$, it
corresponds to a shift $\Delta \Lambda_{\rm EW}=\kappa \Delta
\rho_{\rm vac} \simeq - 0.44~\mbox{cm}^{-2}$ in the cosmological
constant $\Lambda$, while the observed value is given by $\Lambda_{\rm
obs}=\kappa\,\rho_{\rm crit}\,\Omega_{\Lambda}=1.6517
\power{-56}~\mbox{cm}^{-2}$. 
$\Omega_{\Lambda}=0.67^{+0.027}_{-0.023}$ is the dark energy fraction
of the critical energy density $\rho_{\rm
crit}=3\,H_0^2\,\kappa^{-1}\,=1.878\power{-29}\,h^2\,\mbox{gr/cm}^3$
with $h=0.67 \pm 0.02$ for which the universe is
flat~\cite{Ade:2013zuv}.}.  Much later, after the universe has cooled
down to a temparature of about $150~\mv$, QCD undergoes the chiral
phase transition which yields another substantial contributions to the
vacuum density\footnote{The chiral phase transition of QCD which leads
to quark condensates contributing
$$T_{\mu\nu\,\mathrm{QCD}}^\mathrm{vac}=-\vev{\cL_\mathrm{QCD}}\,g_{\mu\nu}=
\left\{m_u\:\bar{u}u+m_d\:\bar{d}d+m_s\:\bar{s}s+\cdots \right\}
\,g_{\mu\nu}$$
to the cosmological constant, has to be reconsidered under the aspect
that the relation between bare an renormalized quantities are physical
in the low energy effective approach. Maybe it is possible to give a
more precise meaning also to the gluon condensate within this
context. The quark condensates yield
$\rho^{\rm vac}_{qq}= m_u\:\bar{u}u+m_d\:\bar{d}d+m_s\:\bar{s}s+\cdots
\simeq -2\times 0.098~\gv^4 - 0.218~\gv^4$ or 
$$\Lambda_{q\bar{q}}\simeq -1.0542\power{-12}~\mathrm{cm}^{-2}\epo$$
The gluon condensate is not well defined. A typical value found in the
literature is $\langle
\frac{\alpha_s}{\pi} G G\rangle \sim (0.389\; \gv)^4 $ which would 
overcompensate the negative quark condensate terms and would change
the result to $$
\Lambda_{\rm QCD} \simeq 1.668 \power{-10} ~\mathrm{cm}^{-2}\epo$$}. 
This shows that in any case the cosmological constant, represented by
the dark energy density, obviously is changing during the
evolution of the universe. As discussed
earlier, the decay of the Higgs inflaton transforms the cosmological
constant, first into top radiation, which gets reduces by $a(t)^{-4}$
until the EW phase transition, and at the EW phase transition into
matter which afterwards diminishes like $a(t)^{-3}$ to date. Thus the
cosmological constant [scaling with $a(t)^{0}$] could have been
decimated on account of other energy forms which get largely diluted
by the expansion of the universe. It is clear that the total energy
density as a function of time
{\small
\bea
\rho(t)=\rho_{0,{\rm crit}}\left\{\Omega_{\Lambda}+\Omega_{0,{\rm k}} \left(a_0/a(t)\right)^2+\Omega_{0,{\rm mat}} \left(a_0/a(t)\right)^3
+\Omega_{0,{\rm rad}} \left(a_0/a(t)\right)^4\right\}
\eea
}
only reflects a present-day snapshot.  The $\Omega$'s representing the
present fraction of dark energy, curvature, matter and radiation of
the total universal Einstein-de Sitter density $\rho_{0,{\rm crit}}$
(globally flat space), do not really describe the evolution of the
energy density in the history of the universe, because physical
processes transform different forms of the energy. Such transmutations
in most cases are well understood and of course have been well
accounted for most of the known processes. However, to my knowledge,
it does not take into account the scenario we advocate, namely that
the universe has undergone the Higgs mechanism after inflation.  In
other words, the cosmological ``fine tuning'' problem could turn out
to be a pseudo problem as the dynamics, subject to energy balance
constraints, resolves itself, in the sense that the cosmological
constant is decaying into other forms of radiation and matter which
naturally decrease with time. The main effect, however, is the vacuum
rearrangement during the Higgs transition as we a going to argue now.

\begin{figure}
\centering
\includegraphics[height=5cm]{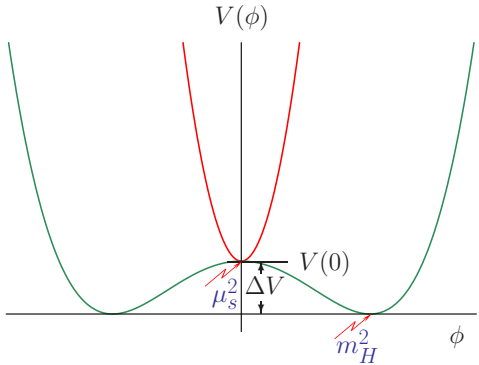}
\caption{Vacuum rearrangement of the Higgs potential. At the zero of
$X(\mu)$ of Eq.~(\ref{finetuning}) $\Delta V=-V(0)$ nullifying the
difference between bare and renormalized CC. The bare mass
$m_0^2=\mu_s^2$ in the symmetric phase is correlated to the renormalized
$m_H^2$ in the Higgs phase, as follows from the text.}
\label{fig:SMHiggsCC} 
\end{figure}

\begin{figure}
\centering
\includegraphics[height=5cm]{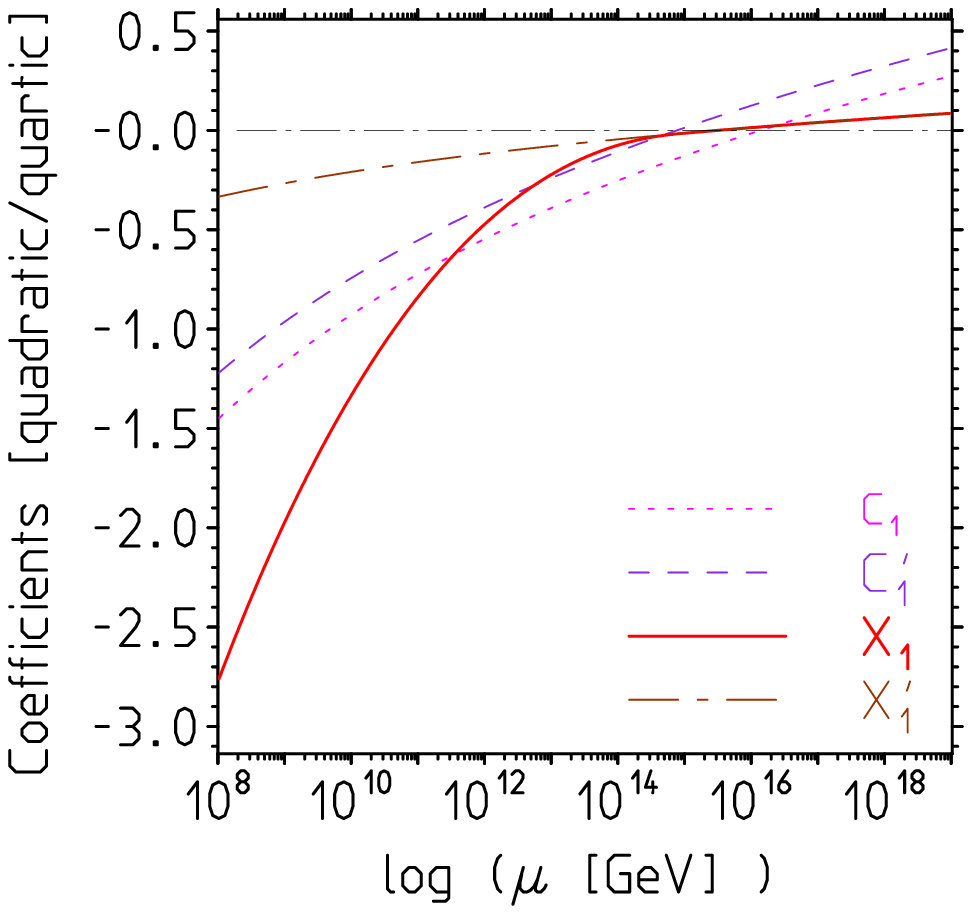}
\includegraphics[height=5cm]{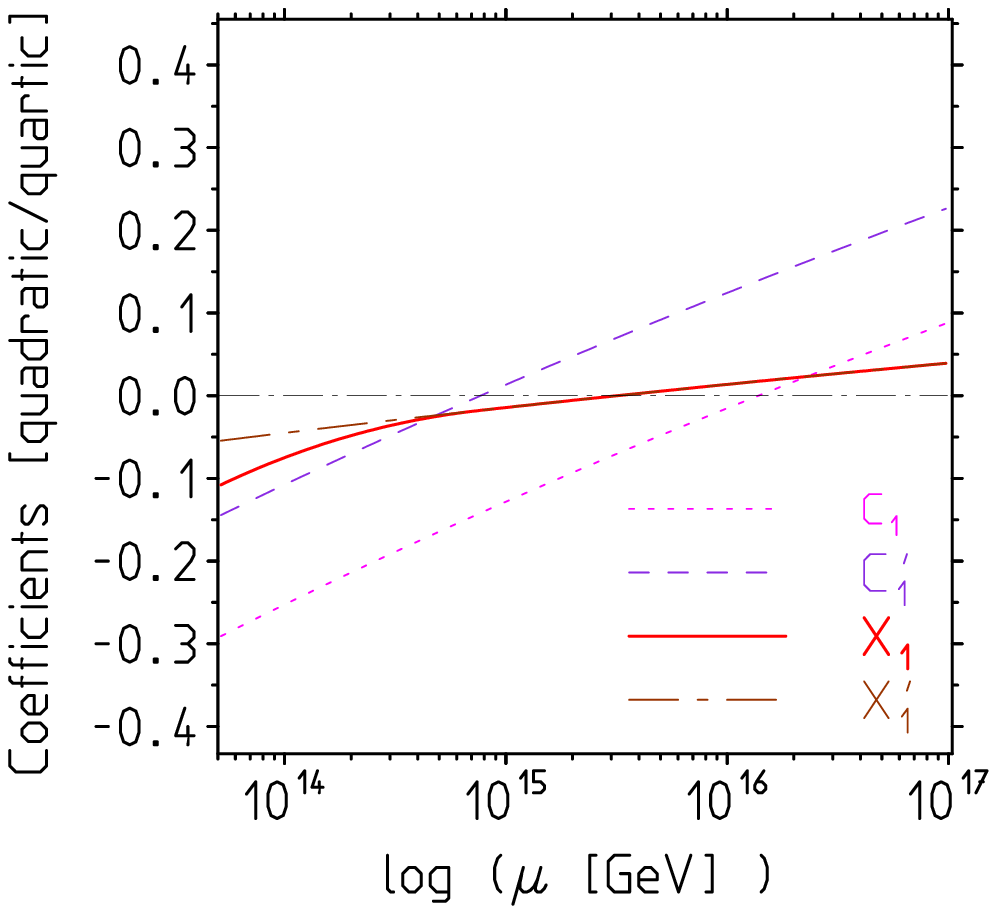}
\caption{The correlated zeros of $C_1(\mu)$ in the quadratic and of
$X_1(\mu)$ in the quartic Planck cutoff enhanced terms,
respectively. Relevant are the effective coefficients
$C'_1=C_1+\lambda$ and $X_1$. Here $X'_1$ represents the quartic
coefficient not including the vacuum rearrangement jump, which does not
actually affect the location of the zero.}
\label{fig:TheZeros} 
\end{figure}

We know that for the early cosmological evolution the CC plays a key
role. The Higgs potential and the Higgs field dynamics are
all-dominant shortly after the big bang and until the Higgs mechanism
is ignited by the sign-flip in our key running function $C(\mu)$
Eq.~(\ref{coefC1}) or $C'(\mu)$ Eq.~(\ref{coefC1prime}). As we have
seen, the Higgs fluctuation field during inflation decays
exponentially and eventually is not able to yield the required blow-up
exponent $N_e$. In fact the predictable constant $V(0)$
Eq.~(\ref{theCC}), provided by quartically ``divergent'' Higgs loops,
supplies the necessary persisting blowing-up of the universe. This
represents the intrinsic CC which depends on $m^2(\mu)$ and
$\lambda(\mu)$ but persists to be large until the bare Higgs mass term
changes sign and the vacuum reorganizes itself (see Fig.~\ref{fig:SMHiggsCC}). From
\bea
V(\phi)=V(0)+\frac{{m'}^2}{2}\,\phi^2+\frac{\lambda}{24}\,\phi^4~~\mathrm{
\ with \ } {m'}^2 <0\,, 
\eea 
we have a minimum at $\phi_0^2=-\frac{6{m'}^2}{\lambda}$ and an effective
bare Higgs mass $m_H^2=-2{m'}^2$. At the minimum $V(\phi_0)=V(0)+\Delta V(\phi_0)$
with $\Delta V(\phi_0)=\frac{{m'}^2}{2}\,\phi_0^2+\frac{\lambda}{24}\,\phi_0^4
=-\frac{3}{2}\frac{{m'}^4}{\lambda}$, and using
${m'}^2=\frac{\mu_m+C+\lambda}{2}\,\Xi$ with $\Xi=\frac{\mpl^2}{16\pi^2}$, we find
\bea
V_{\rm min}&=&V(0)+\Delta V(\phi_0) = \frac{m_0^2}{2}\,\Xi+\frac{\lambda}{8}\Xi^2-\frac{3}{2}\frac{{m'}^4}{\lambda}\nn\\
\hspace*{-8mm}&=& \frac18\,\frac{\mpl^4}{(16\pi^2)^2}\,\biggl(2\,\left(\mu_m+C(\mu)+\lambda(\mu)
\right)-\lambda(\mu)  \nn \\  && 
-\Theta(-(\mu_m+C(\mu)+\lambda(\mu)))\frac{3}{\lambda(\mu)}\,\left(\mu_m+C(\mu)+\lambda(\mu)\right)^2\biggr)\epo
\label{DeltaVjump}
\eea
Here $\Theta(x)$ denotes the step function: $\Theta(x)=1\,,x>0$ and $=0\,,\,\,x<0\,.$
The subtraction of the ``jump'' $\Delta V(\phi_0)$ applies as soon as
${m'}^2 <0$, equivalently, $\mu_m+C(\mu)+\lambda(\mu)<0\cs$ which is
in the Higgs phase.  Here again we observe an intriguing structure,
which exhibits a zero not far away from the zero of $C(\mu)$ as we
have $\lambda>0$ but small and $C(\mu)$ is growing negative below its
zero. So actually, the effective CC counterterm has a zero, which again is a point
where renormalized and bare quantities are in agreement:
\bea
\rho_{\Lambda 0}=\rho_{\Lambda} +\frac{\mpl^4}{(16\pi^2)^2}\,X(\mu) 
\label{finetuning}
\eea
with $X(\mu)=0$ close to the zero of $C(\mu)$.  The impact on the
various terms in the potential is displayed in
Fig.~\ref{fig:Lagrangianext}. Interestingly, the functions $C(\mu)$
and $X(\mu)$, which nullify the difference between renormalized and
bare mass and vacuum density, respectively, are strongly correlated,
implying that the corresponding zeros are effective at comparable
scales. It means that short range and long range regime match in a
vicinity of the Higgs transition spot. One thing is clear, there is no
dramatic fine tuning problem as anticipated usually. The mechanisms
both for the quadratic- as well as for the quartic-enhancements are
not a matter of taking differences between two independent huge
numbers, but a matter of a huge number which is multiplied by a
function exhibiting a zero by cancellation of normal sized effective couplings
in our cases. That is how self-organized fine-tuning works. What is
also interesting is that these mechanisms are possible only by
cancellations between bosonic and fermionic contributions. 
In Fig.~\ref{fig:TheZeros} we show the location and structure of the
correlated zeros. The sign-change for the bare mass $m_0^2=m^2+\delta
m^2$ occurs at the zero of $\mu_m+C(\mu)$, where $\mu_m$ accounts
for a small positive possible contribution from a renormalized mass
(in Plank mass units). In fact, because of the necessary Wick
reordering to account for the vacuum contributions the true effective
mass is $m_0^2+\frac{\lambda}{2}\,\Xi$ such that the effective Higgs
transition is ruled by $\mu_m+C(\mu)+\lambda(\mu)$ not by $\mu_m+C(\mu)$,
as we see in Fig.~\ref{fig:TheZeros} this shifts the location of the
Higgs transition to a lower value $\mu'_0\simeq 7.7\power{14}~\gv$!
Again, we point out that such results sensitively depend on the
\MSb input parameter values, and are expected to change if more
precise parameter values and a better understanding (including higher
order effects) of the matching relations have been achieved. Above the
zero of ${m'}^2$ the coefficient of the ``amplifier''
$\frac{\mpl^4}{(16\pi^2)^2}$ is
$2\,(\mu_m+C(\mu)+\lambda(\mu))-\lambda(\mu)>0\,,$ which induces a huge
CC. As $\mu_m+C(\mu)+\lambda(\mu)$ flips sign we get a negative jump
$-\frac32\,\frac{(\mu_m+C(\mu)+\lambda(\mu))^2}{\lambda(\mu)}$. Surprisingly,
at the zero of ${m'}^2$ the quartic coefficient $X(\mu)=-\lambda$,
i.e. $X(\mu)$ changes sign slightly before the Higgs transition at
$\mu_{\rm CC}\simeq 5.01\power{15}~\gv$! Thus, again it is the running
of the couplings and not the jump from the vacuum rearrangement which
is responsible for attaining the zero, which is the matching point
between bare and renormalized quantities!

\begin{figure}
\centering
\includegraphics[height=6cm]{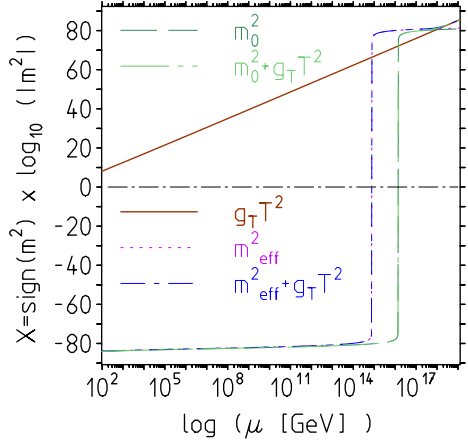}
\caption{The effective mass vs. the bare mass including finite
temperature effects. The latter have little effect on the location of
the EW phase transition. The shift $C_1 \to C'_1=C_1+\lambda$ has a
substantial effect on the scale of the Higgs transition.}
\label{fig:lambdashiftFT} 
\end{figure}
In order to understand better what I mean by ``self-organized
fine-tuning'' let us consider the lattice version of the SM. This is
an obvious candidate for our Planck medium in the right universality
class by construction and it is a non-perturbatively well-defied
system.  If we investigate this lattice SM at short distances we see
the lattice structure as the true short distance world and we would
see it to be in the symmetric phase and a relation like (\ref{barem2})
or (\ref{finetuning}) to have a true physical meaning. In this short
distance regime the bare system is the physical one as we have been
anticipating in our description of the inflation era. In contrast, if
we investigate its long range properties by probing appropriate
observable quantities, we would see the effective renormalizable
continuum field theory and the symmetry to be spontaneously
broken. This requires parameters of the lattice system to be in a
range which allows that long range order can be effective. As
elaborated in Sect. 2 of Ref.~\cite{Jegerlehner:2013cta}, before we
renormalize the long range correlations emerging from the lattice
system, these indeed exhibit a residual dependence on the
cutoff. However, the cutoff at long distances functions as a
renormalization reference scale only and has lost its meaning as a
cutoff\footnote{Formal criterion is that the RG in the cutoff
$\Lambda$ takes a homogeneous form.}! This means that the cutoff can
be renormalized away in favor of the \MSb parameter $\mu$, for
example. By low energy instruments we will no longer be able to probe
the short distance structure, therefore we will not encounter any
cutoff related effect and everything plays in the framework of the
renormalized low energy effective theory. But, how do we get ride of
the cutoff in the mass renormalization? Indeed we have to tune the
bare mass of the short distance system to criticality, i.e., tune
$m_0^2\to m^2_{0\,{\rm crit}}$ such that $m^2=+0$. The renormalized
mass now given by $m^2=m_0^2-m^2_{0\,{\rm crit}}$ then can be tuned to
have any value we want. This is fine tuning par excellence, at least
in the symmetric phase. However, this argument does not answer the
question when we are in the spontaneously broken phase where long
range order is taking over. I think that the hot Planck medium is
exhibiting a plethora of modes where some conspiring ones are able to
reach to long distances, exhibiting the masses we see. Why the minimum
of the Higgs potential should not naturally be close to the one at
zero of the symmetric phase? Why should it jump from zero to $\mpl$
suddenly? And if the location of the minimum
$\phi_0 \ll \mpl$ is natural, why $m_H \ll \mpl$ is not? Our
calculation presented above proves that as the universe expands the
spontaneous symmetry breaking phase develops continuously from the
symmetric phase, i.e. as ${m'}^2$ passes a zero
$\phi^2_0(\mu)=-\frac{6{m'}^2}{\lambda}(\mu)$ moves smoothly from zero
to whatever value, and in any case at some point matches its
renormalized value. No low energy experiment is able to substantiate a
supposed fine-tuning problem, and doing short distance experiments
would probe the symmetric phase where a fine-tuning problem is not
known to exist, at least as long as we do not know what the value of
the renormalized mass in the symmetric phase is.
 
\begin{figure}
\centering
\includegraphics[height=6cm]{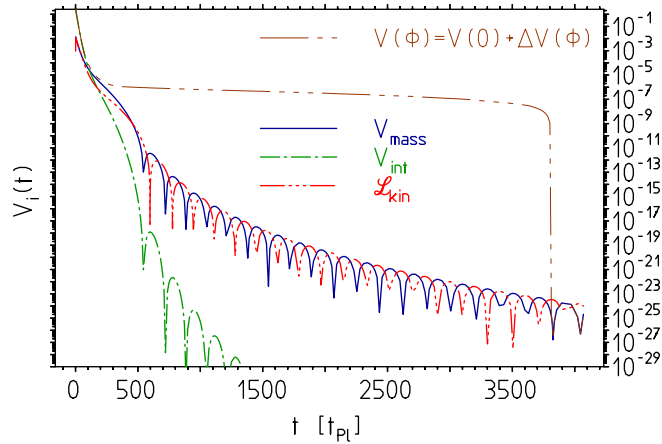}
\caption{Here we show the same quantities as in
Fig.~\ref{fig:Lagrangian} in an extended range exhibiting the vacuum
rearrangement at $\mu_{\rm CC}$. After inflation the scene is
characterized by a free damped harmonic oscillator behavior. Relevant
scales are $\mu_0\simeq 1.4\power{16} \Leftrightarrow t\simeq 870
\mpl^{-1}$ at the zero of $m^2_0-m^2=0$, $\mu_{\rm
CC}\simeq 3.1 \power{15} \Leftrightarrow t\simeq 4000 \mpl^{-1}$ where
$\rho_{\Lambda 0}=\rho_{\Lambda }$ and
$\mu'_0\simeq 7.7\power{14}\Leftrightarrow t\simeq 15844 \mpl^{-1}$
the true Higgs transition point ${m'}^2=0$.}
\label{fig:Lagrangianext} 
\end{figure}
The dynamical part now is the broken phase Higgs Lagrangian
\bea
\cL_{\rm Higgs}=\frac12 \dot{\phi}^2+\frac{m_H^2}{2}\,\phi^2+ \frac{\sqrt{3\lambda}m_H}{6}\,\phi^3+\frac{\lambda}{24}\,\phi^4\epo
\eea
where $m_H=m_H(\mu)$ is the renormalized \MSb mass. The CC is also to
be identified with some renormalized value, which we know must be
small. The field equation (\ref{fieldeq}) now reads
\bea
\ddot{\phi}+3H\dot{\phi}=-\left(m_H^2\,\phi+ \frac{\sqrt{3\lambda}\,m_H}{2}\,\phi^2+\frac{\lambda}{6}\,\phi^3\right)\epo
\label{brokenfieldeq}
\eea
I expect that the observed value of dark energy has to be considered
as a phenomenological constraint. The reason is that $\rho_\Lambda$ is
dependent on the Higgs field magnitude, which is not fixed by other
observations, except maybe by CMB inflation data. In addition we have
to keep in mind that our scenario is very sensitive to our basic
parameters $C(\mu)$ and $\lambda(\mu)$, which were obtained by evolving
coupling parameters over 16 orders of magnitude in scale. This cries for high
precision physics to really settle the issue. Note however, that given
the SM couplings everything here is a SM prediction without any extra
assumption. In any case, what we learn is that the quartic divergences
in the vacuum energy are not the problem, rather they are the solution
providing inflation as a necessary consequence of the SM.

What about the other vacuum condensate contributions we expect to show
up in the broken phase of the SM? First the Higgs transition
contribution associated with a non-vanishing $\braket{H}=v$ now is
accounted for by the ``jump'' $\Delta V(\phi_0)$ contributing to
Eq.~(\ref{DeltaVjump}) and thus has been accounted for in the
relation (\ref{finetuning}). What concerns the QCD condensates
contribution mentioned earlier, this has to be reconsidered in our
context of the LEESM. Presently, not even the sign of $\rho^{\rm vac}_{\rm
QCD}$ is known for sure, so there is a chance that also this problem
will find its solution.

\section{Conclusions}
The recent discovery of the Higgs at the LHC revealed a Higgs mass in
a window which strengthens our believe into a low energy effective SM
scenario, with a largely unknown medium at the Planck scale exhibiting
the Planck mass as a cutoff. This may represents a dramatic change of
the predominating paradigm concerning the ``Path to physics at the
Planck scale'' which is the SUSY, GUT and strings driven `` the higher
the energy the more symmetry'' belief. The LEESM scenario supports
strongly an emergent look at what we see at long distances. The SM is
a naturally emergent low energy structure, the world as seen from
far away~\cite{Jegerlehner:2013cta}.

The Higgs discovery, together with the fact than non-SM physics has
not yet shown up at the LHC, may have a dramatic impact on particle
theory and particle physics all-together. We have shown that, under the
conditions that the SM vacuum remains stable up to the Planck scale
and that the quadratically enhanced Higgs mass counterterm exhibits a
zero not far below the Planck scale, likely the SM largely summarizes
the all-driving laws of physics which govern the evolution of the
universe from its birth and possibly for all future. This likely
brings high precision physics and high precision SM calculations in
the focus of future developments as a tool to learn more about early
cosmology. We note that close-by non-SM low energy emergent new
physics is naturally expected to exist.  The origin of dark energy
or the strong CP problem definitely may find their solution
in new not yet fully uncovered low energy emergent physics~\cite{Jegerlehner:2013cta}.
Should vacuum stability in the SM fail all this could be completely 
different~\cite{Bian:2013jra,Masina:2013wja,Hamada:2013mya}.

What we have shown is that a CMB data consistent inflation scenario is
possible solely on the basis of SM physics, with the Higgs being the
driver.  The big difference in comparison to alternative inflation
scenarios is the fact that we almost perfectly know the properties of
the inflaton, such that we are able to get true predictions, results
which are more than more or less direct consequences of more or less
plausible assumptions.

We essentially are left with two quantities which we have to constrain
by data extracted from the observed CMB fluctuation patterns: the
renormalized mass in the symmetric phase of the SM, and more
importantly, the magnitude of the classical Higgs field at the Planck
scale.  The renormalized mass square we assumed to be subleading at
the Planck scale such that $m^2 \ll m^2_0$ at
$\mpl$. This seems to be well supported by CMB data. The second
assumption derives from the need of sufficient inflation, required to
solve the CMB horizon problem, in particular. We found that a 25\%
enhancement of a field strength $\phi$ which yields an energy density
$\rho_\phi\simeq V(\phi)\simeq \mpl^4$ is sufficient to provide the
necessary magnitude of exponential growth of the FRW-radius. We
include the Higgs vacuum diagrams as predicted by the SM and which
provides a very weakly decreasing moderately big cosmological constant
contribution $V(0)$, which depends only on the running values of
$\lambda(\mu)$ and $C(\mu)$, the coefficient function which determines
the enhanced effective Higgs mass. What at first sight looks to be a
severe cosmological constant problem, resolves itself, as the
difference between the renormalized and the bare cosmological constant
vanishes near slightly above the Higgs transition point by running of
the SM couplings also in this case. At this point bare and
renormalized values of the cosmological constant coincide and the
renormalized value may be identified with the observed tiny dark
energy term, which in spite of its smallness remains the dominant
contribution of today's energy density in the universe, as we
know. This does not exclude the possibility that a better
understanding of the dynamics of the EW phase transition would allow
us to predict $\rho_\Lambda$. Likely, also today's value of the dark
energy is provided by the Higgs as the source which is continuously
blowing energy into our universe providing the accelerated
expansion. This corresponds to a tiny continuous
inflation. Remarkably this does not contradict energy conservation as
the cosmological constant is the only covariant quantity which is
covariantly conserved by-itself.

This also sheds new light
on the hierarchy problem, usually considered to be a fine-tuning
problem in the sense that a relatively small physical quantity is the
difference of two uncorrelated huge numbers. In fact the huge cutoff
terms turn out to be multiplied by an $O(1)$ function which can
exhibit a zero, which actually at some point removes the cutoff dependence
and  provides a spot at which the bare short distance world matches with
our renormalized long distance world.

What is new here is that we have evidence that the disentanglement
between short distance and long distance is complete. Quadratically as
well as quartically blown-up quantities, natural in the bare system,
decouple from low energy physics at the Higgs transition spot where
long range order in form of the Higgs vacuum condensate $v$ sets the
scale.  Now long range effects shape what we are able to see.  The
large hierarchy $v/\mpl$ just tells us that our world is close below a
second order phase transition point\footnote{I am referring here to
the commonly known example of spontaneous magnetization in a
ferromagnetic system: the magnetization $M$ is the order parameter
(corresponding to our $v$ in the SM), the result of long range
collective behavior of the spins of the system. Above a critical
temperature $T_c$ there is no spontaneous magnetization $M(T)\equiv0$
as $T>T_c$. Below $T_c$, as we lower $T$, $M(T)$ is a monotonically
increasing function with its maximum value at $T=0$. If we approach
$T_c$ from below $M(T)$ continuously decreases to zero at $T_c$, and
hence can be arbitrarily small depending on how close we are to the
critical point $(T,M)=(T_c,0)\,,$ which corresponds to a second order
phase transition point, the end point of a continuous family of first
order phase transitions corresponding to possible jumps in the
magnetization $M(T)\leftrightarrow -M(T)$ when $T<T_c$.}. This is also
what the theory of critical phenomena and emergent continuum field
theory structures suggest.

One more remark should be made here: as I pointed out several times
the spot in SM parameter space where SM inflation can work seems to be
very narrow.  The detailed SM inflation scenario, e.g. what are the
predominant characteristics as a function of time in the evolution of
the early universe, depends a lot on the precise value of the \MSb top
Yukawa coupling $y_t(M_Z)$ in particular. Thus details may change a lot
when our knowledge of the parameters improve. Nevertheless, I think
that this analysis raises hopes that at the end we will be able to 
establish the Higgs as the inflaton and as the supplier of dark energy.

\bigskip

\noindent
\newpage

\section*{Acknowledgments}

Many thanks to Paul S\"oding for critically reading the manuscript and
his helpful comments and questions. I furthermore thank Oliver B\"ar, Nigel
Glover, Mikhail Kalmykov and Daniel Wyler for their interest and for
many clarifying discussions. I gratefully acknowledge the interest
and patience of my wife Marianne, as many of the ideas worked out here
came to my mind when I tried to explain her what I think the role the
Higgs could have played in the evolution of the early universe.

\end{document}